\documentclass[11pt]{article}

\usepackage[utf8]{inputenc}
\usepackage{amsmath,amsfonts}
\usepackage{color,enumerate}
\usepackage{graphics,afterpage,subfigure,graphicx,subfigure,wrapfig,lipsum,caption}
\usepackage[utf8]{inputenc}
\usepackage{mathtools}   
\usepackage{pifont}
\usepackage{natbib}
\usepackage[OT1]{fontenc}
\usepackage{amscd}
\usepackage{latexsym,amssymb,amsthm}
\usepackage{pdflscape}
\usepackage{hyperref}
\usepackage{booktabs}
\usepackage{threeparttable}

\newcommand{\RR}{\mathbb R}
\newcommand{\bfw}{{\bf w}}
\newcommand{\bfx}{{\bf x}}
\newcommand{\bfz}{{\bf z}}
\newcommand{\bfy}{{\bf y}}
\newcommand{\bfc}{{\bf c}}
\newcommand{\bfu}{{\bf u}}

\newcommand{\bfro}{\boldsymbol{\rho}}

\newcommand{\bta}{\boldsymbol{\theta}}
\newcommand{\bfmu}{\boldsymbol{\mu}}
\newcommand{\bfpi}{\boldsymbol{\pi}}

\newtheorem{lemma}{Lemma}
\newtheorem{remark}{Remark}
\newtheorem{theorem}{Theorem}

\begin{document}

	\title{\mbox{}\\[-11ex]Estimating densities with nonlinear support using Fisher-Gaussian kernels}
	\vspace{-5ex}
	\author{\\[1ex]Minerva Mukhopadhyay$^{1}$\thanks{The authors contributed  equally to this paper.}, Didong Li$^{2}$\footnotemark[1],  and David B Dunson$^{2,3}$ \\ 
		{\em Department of Mathematics and Statistics, Indian Institute of Technology, Kanpur$^1$,} \\ {\em Department of Mathematics, Duke University$^2$,} \\{\em Department of Statistical Science, Duke University$^3$}}
	\date{\vspace{-5ex}}
	
	\maketitle
			Current tools for multivariate density estimation struggle when the density is concentrated near a nonlinear subspace or manifold. Most approaches require choice of a kernel, with the multivariate Gaussian by far the most commonly used. Although heavy-tailed and skewed extensions have been proposed, such kernels cannot capture curvature in the support of the data. This leads to poor performance unless the sample size is very large relative to the dimension of the data. This article proposes a novel generalization of the Gaussian distribution, which includes an additional curvature parameter. We refer to the proposed class as Fisher-Gaussian (FG) kernels, since they arise by sampling from a von Mises-Fisher density on the sphere and adding Gaussian noise. The FG density has an analytic form, and is amenable to straightforward implementation within Bayesian mixture models using Markov chain Monte Carlo. We provide theory on large support, and illustrate gains relative to competitors in simulated and real data applications.
	
	
	\section{Introduction}

Density estimation is one of the canonical problems in statistics and machine learning.  Even when the focus is not directly on the density of the data, density estimation often is a key component of the analysis, arising in clustering, classification, dimensionality reduction, and robust modeling applications among many others. Our focus is on kernel-based approaches to multivariate density estimation, with an emphasis on mixture models.   In the simplest setting, one has observed data $\bfx_i = \left(x_{i,1},\ldots,x_{i,d}\right)’,\ i=1,\ldots,n$, which can be assumed to be independent and identically distributed (i.i.d.) draws from an unknown density $f$ that can be expressed as: 
\begin{eqnarray}
f(\bfx) = \sum_{k=1}^M \pi_k \mathcal{K}\left(\bfx; \bta_k\right),
\end{eqnarray} \label{eq:mixmodel}
where $\bfpi = \left(\pi_1,\ldots,\pi_M\right)’ $ is a vector of probability weights summing to one, $M$ is the number of mixture components, and $\mathcal{K}\left(\bfx ;\bta\right)$ is a kernel having parameters $\bta$.  

Model (\ref{eq:mixmodel}) is extremely widely used, forming the foundation of model-based clustering, model-based approaches to density estimation and associated non/semiparametric modeling, and Bayesian nonparametrics.  There is an immense literature focused on different approaches for choosing the weights $\bfpi$ and number of components, see \cite{fruhwirth2006finite} and \cite{ghosal2017fundamentals}.  It is also well known that the choice of kernel $\mathcal{K}(\bfx;\bta)$ is very important, not only in applications in which one wants to use the model for clustering and cluster-based inferences (see, e.g., \cite{miller2018robust}) but also for accurate density estimation.  However, even with this rich literature, the class of kernels that can be used practically in inferences under (\ref{eq:mixmodel}) remains quite restrictive from the perspective of characterizing {\em curvature} in the data.

\begin{figure}[h]
	\begin{center}
		\includegraphics[scale=.5]{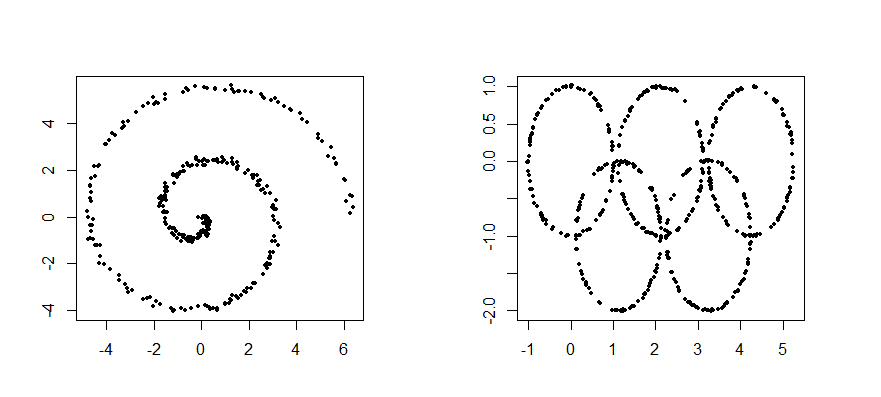}  
		\caption{Scatter plot of 300 points from noised spiral (left panel), and $500$ points from Olympic rings (right panel) }\label{fig_intro1}
	\end{center}
\end{figure}

We clarify the problem using two toy examples - the noised spiral and Olympic rings - shown in Figure \ref{fig_intro1}.  Both are examples of $d=2$ dimensional data concentrated near a one dimensional curve(s).

These are common toy examples from the literature on nonlinear dimensionality reduction and manifold learning (\cite{lee2007nonlinear}).  We use them as motivation, since even in these simple cases problems arise for usual Gaussian kernels due to their intrinsic linear structure.  In general for real world datasets, we will have limited prior knowledge about the support of the data, but in general no reason to suspect that it should be close to linear even locally.  Indeed, in application areas ranging from imaging to signal processing to astrophysics, there is often strong reason to suggest substantial curvature in the data.	

\begin{figure}
	\begin{center}
		\includegraphics[height=1.45 in, width=5.5 in]{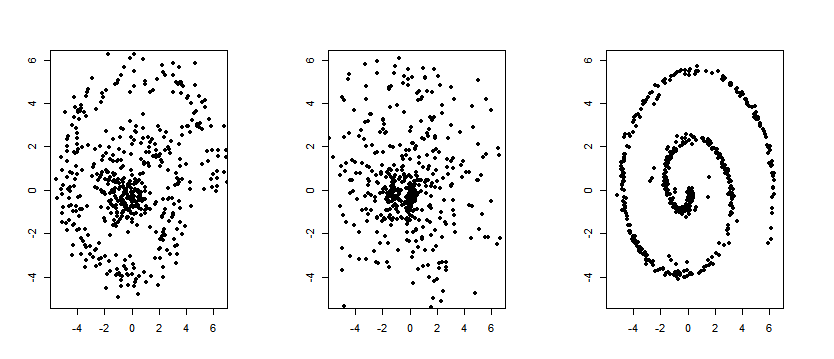} \includegraphics[height=1.45 in, width=5.5 in]{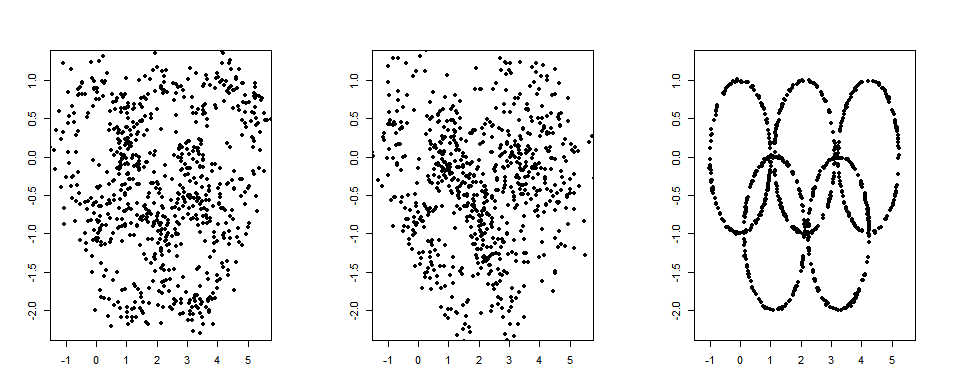}
		\caption{Scatter plot of sample points from predictive densities obtained by KDE (left), DP mixture of Gaussian (middle) and Fisher-Gaussian kernel mixture (right). The first and second rows correspond to the left and right panels from Figure \ref{fig_intro1} }\label{fig_intro2}
	\end{center}
\end{figure}

Figure \ref{fig_intro2} shows results from estimating the density of the data shown in Figure \ref{fig_intro1} using frequentist kernel density estimation (KDE, see, e.g., \cite{Silverman_book}) implemented in the `ks' R-package, a Bayesian nonparametric approach relying on Dirichlet process mixtures (DPM, see, e.g., \cite{maceachern1998}) of Gaussian kernels implemented in the `dirichletprocess' R-package, and our proposed approach, which uses (\ref{eq:mixmodel}) within a Bayesian inference approach incorporating a novel class of kernels that accommodate curvature.  For easy visualization, we plot new samples from the estimated density in each case.  It is clear that the KDE and DPM of Gaussian methods fail to accurately estimate the density.  This is due in large part to the fact that the multivariate Gaussian kernel has elliptical contours about a line or plane, leading to challenges in accurately locally approximating the densities of the data shown in Figure \ref{fig_intro1}.  The Gaussian kernel provides an accurate local approximation only in a very small region, and hence one needs many kernels and a correspondingly large sample size $n$ for good performance.

A natural question is whether this problem can be solved with the various alternative kernels that have been proposed in the literature.  In the multivariate case, the most popular extensions have been to inflate the tails using multivariate-t kernels (\cite{McLachlan1998}, \cite{Lee2016}) and/or to generalize the Gaussian distribution to allow skewness (\cite{Azzalini1996}, \cite{Azzalini2008}).  Motivated by the problem of more robust clustering, there are also methods available for nonparametrically estimating the kernels subject to unimodality (\cite{rodriguez2014univariate}) and log concavity (\cite{hu2016maximum}) restrictions.  Also to improve robustness of clustering, one can use a mixture of mixtures model employing multiple kernels having similar location parameters to characterize the data within a cluster (\cite{malsiner2017identifying}).  None of these methods address the fundamental problem we focus on, which is how to better characterize curvature in the support.

An alternative strategy would be to abandon kernel-based methods entirely in attempting to characterize the density of $f$ more accurately.  In this regard, one could potentially use flexible non-linear latent structure models developed in the machine learning literature, ranging from Gaussian process latent variable models (GP-LVMs) (\cite{li2016review}, \cite{doersch2016tutorial}) to variational auto-encoders (VAEs) leveraging deep neural networks (\cite{lawrence2005probabilistic}).  However, such methods are complex black boxes that fail to produce an analytic expression for the density of the data and sacrifice the simple interpretation of (\ref{eq:mixmodel}) in terms of producing data clusters that each have a known distributional form.  

With this motivation, we propose to generalize the Gaussian kernel to accommodate curvature in a simple and analytically tractable manner.  In particular, we develop a new class of Fisher-Gaussian (FG) kernels by generating observations from a von Mises-Fisher density on a sphere and adding Gaussian noise.  The resulting FG distribution has an analytic form, and includes a curvature parameter, in addition to location and scale parameters.  Using this kernel, we propose FG mixture models and implement these models within a Bayesian framework using Markov chain Monte Carlo (MCMC).

Section 2 derives the FG class of kernels and studies basic distributional properties and representations.  Section 3 includes these kernels within a Bayesian mixture model, and develops a straightforward MCMC algorithm for posterior computation. Section 4 studies asymptotic properties.  Section 5 contains a simulation study. Section 6 applies the methods to several datasets, and Section 7 contains a discussion. Proofs are included in the Appendix.

\section{The Fisher-Gaussian kernel}\label{sec:1}
Suppose the observed data, $\bfx_1,\ldots, \bfx_n$, are supported around a sphere with center $\bfc$ and radius $r$. Due to noise we do not observe points exactly on the sphere. Therefore we model the data as
\begin{equation}\label{eqn:model}
\bfx_i =\bfc+ r\bfy_i + \boldsymbol{\epsilon}_i \quad \mbox{where }~~ \boldsymbol{\epsilon}_i \sim N\left({\bf 0}, \sigma^2 I\right), 
\end{equation}
for some unknown noise variance $\sigma^2>0$. Here $\bfc+ r\bfy_i$ is the part of $\bfx_i$ supported exactly on the sphere with center $\bfc$ and radius $r$, and $\boldsymbol{\epsilon}_i$ is a random noise. The random variable $\bfy_i=\left(\bfx_i-\bfc-\boldsymbol{\epsilon}_i\right)/r$ has support on the unit sphere, and is assumed to follow a von-Mises Fisher density as $\bfy_i \sim f_{\mathrm{vMF}}\left(\cdot \mid \bfmu, \tau\right)$, where $\bfmu$ is the mean direction, and $\tau$ is a concentration parameter. Higher values of $\tau$ indicate higher concentration of the distribution around $\bfmu$. Thus the density of $\bfx$ is
\begin{align}
f(\bfx)&= \int_{\RR^d} \frac{1}{r^d} f_{\mathrm{vMF}}\left\{\frac{1}{r} \left(\bfx-\bfc - \boldsymbol{\epsilon}\right) \mid \bfmu,\tau \right\} \phi_{\sigma^2} \left(\boldsymbol{\epsilon} \right) d \boldsymbol{\epsilon} \notag\\
&\hspace{1.5 in}=\int_{\RR^d} \frac{1}{r^d}  f_{\mathrm{vMF}}\left\{ \frac{1}{r} \left(\bfz-\bfc \right) \mid \bfmu,\tau  \right\} \phi_{\sigma^2} \left(\bfx-\bfz \right) d \boldsymbol{\bfz},\label{eqn:FGkernel}
\end{align}
where $\phi_{\sigma^2}(\cdot)$ is the multivariate normal density with mean ${\bf 0}$ and variance $\sigma^2I$.
We refer to (\ref{eqn:FGkernel}) as the Fisher-Gaussian (FG) density. 

\begin{figure}
	\begin{center}
		\includegraphics[scale=.3,trim={2cm 0 3cm 0},clip]{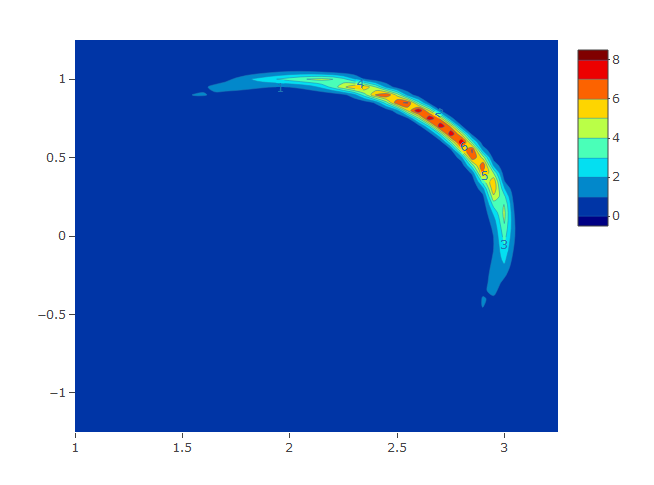} \quad \includegraphics[scale=.3,trim={2cm 0 3cm 0},clip]{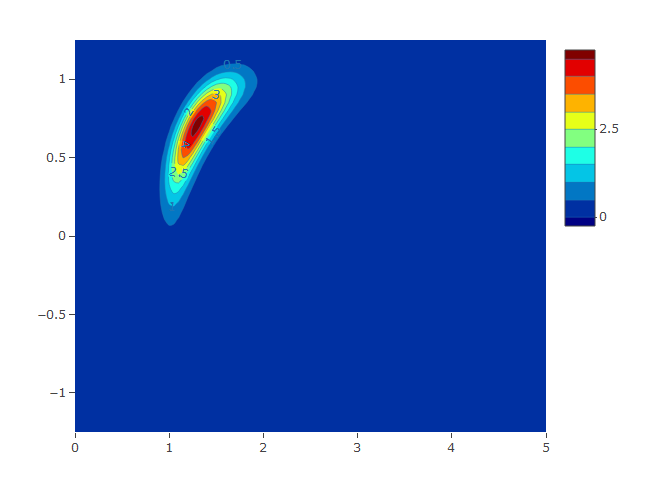}	\quad \includegraphics[height=3.9 cm, width=4 cm,trim={4cm 0 5.1cm 0},clip]{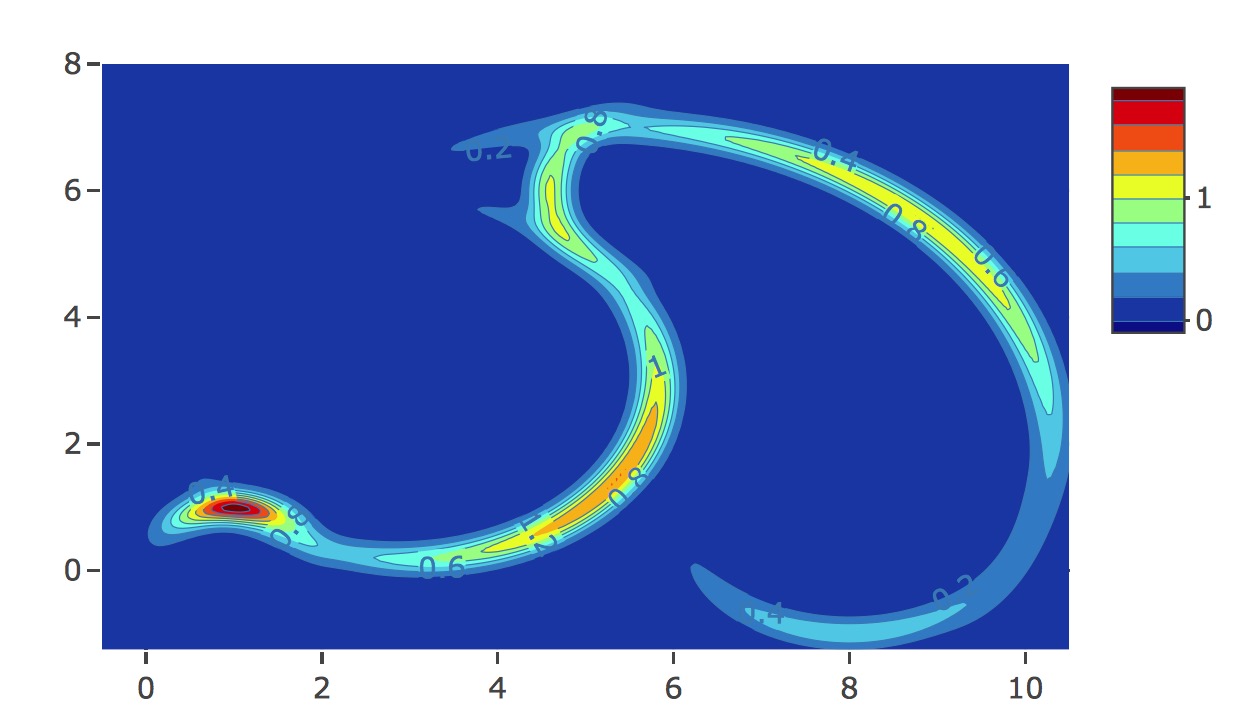} 	
		\caption{Fisher-Gaussian kernel for different parameters (a) (left) $\bfmu=\left(1/\sqrt{2},1/\sqrt{2}\right)$, $\tau=3$,  $\sigma^2=.001$, $\bfc={\bf 0}$, $r=1$; (b) (middle) $\bfmu=\left(-1/\sqrt{2},1/\sqrt{2}\right)$, $\tau=10$, $\sigma^2=.01$, $\bfc=(2,0)$, $r=1$; (c) (right) A typical example of FG-kernel mixture density.  }\label{fig1}
	\end{center}
\end{figure}

As the definition suggests, the FG density is essentially a Gaussian density concentrated near the circumference of a sphere. The kernel depends on five parameters: $\bfc$ and $r$: the center and radius of the sphere along which the kernel is distributed; $\bfmu$ and $\tau$: the location on the sphere around which the density has highest concentration and the precision along the sphere; and $\sigma^2$ controls how far from the sphere data points concentrate. Figure \ref{fig1} (a)--(b) demonstrate the effect of the different parameters of the kernel. Throughout this paper, we denote the FG kernel as $\mathrm{FG}_{\sigma}\left(\cdot\mid \bfro\right)$ where $\bfro=\left(\bfc,r,\bfmu,\tau\right)$. 

The FG-kernel has a simple analytic form which makes it theoretically tractable, and easily applicable. By integrating out $\bfz$ in (\ref{eqn:FGkernel}), the FG kernel can be expressed as follows
\begin{eqnarray}
\mathrm{FG}_{\sigma}(\bfx\mid\bfro)=  \frac{C_d(\tau) \left(2\pi\sigma^2\right)^{-(d/2)}}{C_d\left(\|\tau \bfmu +r(\bfx-\bfc)/\sigma^2\|\right)} 
\exp\left\{ -\frac{1}{2\sigma^2} \left(\|\bfx-\bfc\|^2+r^2\right)\right\}, \label{FG}
\end{eqnarray}
where $C_d(\tau)=\tau^{d/2-1}(2\pi)^{-d/2} I_{d/2-1}^{-1}(\tau)$ and $I_{\nu}$ denotes the modified Bessel function of the first kind of order $\nu$. 
The following lemma proves the above claim.
\begin{lemma}\label{lm:1}
	\begin{enumerate}[a.]
		\item The density in (\ref{eqn:FGkernel}) can be written as in (\ref{FG}).
		\item Consequently, for any given $\bfro$ and $\sigma^2$, $\int_{\RR^d} \mathrm{FG}_\sigma\left(\bfx \mid \bfro\right) d\bfx =1$.
	\end{enumerate}
\end{lemma}

The FG kernel provides a simple and flexible generalization of the spherical Gaussian density, commonly used in multivariate density estimation, to allow ``curved'' densities, as illustrated in Figure \ref{fig1}. The curvature can be modified arbitrarily by changing the radius $r$, with smaller $r$ providing larger curvature and larger $r$ providing densities close to Gaussian.  Potentially, one can define a broader class of FG distributions by replacing either or both of the scale parameters $\tau$ and $\sigma^2$ with positive semidefinite matrices, obtaining a flexible class of curved multivariate distributions. However, we leave this to future work.

\section{Mixtures of Fisher-Gaussian Kernels} \label{sec:2}
Although the FG distribution is useful in other settings as a simple but flexible generalization of the spherical Gaussian density, our focus is on using FG kernels in mixture models. Real data often involve nonlinear relationships among variables, and in such cases mixtures of FG kernels are expected to provide more parsimonious density approximations requiring many fewer mixture components to obtain an accurate approximation. 

To make the model more parsimonious, we propose a nested modification to (\ref{eq:mixmodel}). We start by letting $\bfx_i=\bfz_i+\boldsymbol{\epsilon}_i$, as in model (\ref{eqn:model}), with $\bfz_i$ falling exactly on a sphere. The distribution of $\bfz_i$ depends on parameters controlling inter ($\bfc$ and $r$) and intra-sphere ($\bfmu$ and $\tau$) characteristics. The parameters $(\bfc,r)$ control the location and size of the sphere on which $\bfz_i$ lies,  whereas $(\bfmu,\tau)$ controls the modal point and precision of the distribution of $\bfz_i$ on the sphere. Within a sphere, the distribution can be unimodal or multimodal.  Hence, it is natural to consider a model in which one has a collection of spheres, having varying centers and radii, along with a collection of kernels on each of these spheres.  Thus, we can represent the distribution of $\bfz_i$ as 
\begin{eqnarray*}
	\frac{1}{r_{s_i}} \left(\bfz_i - \bfc_{s_i} \right) \sim f_{\mathrm{vMF}}\left(\cdot \mid \bfmu_{k_i}^{s_i}, \tau_{k_i}^{s_i} \right), \label{eq:fgmix1}
\end{eqnarray*}
where $s_i$ indexes the sphere that $\bfz_i$ lies on, and $k_i$ indexes the kernel within the sphere from which $\bfz_i$ is drawn. The number of spheres, as well as vMF kernels is not known. To keep the representation simple, yet flexible, we let $s_i \in \{ 1,\ldots, \infty\}$ and $k_i \in \{1,\ldots, M\}$. Under the proposed specification, we  have $P\left(s_i = l\right) =\lambda_l$ for $l=1,\ldots,\infty$, and  $P\left(k_i = k \mid s_i=l\right) = \pi_{k}^l$ for $k=1,\ldots,M$.  

Conditionally on $\left(s_i,k_i\right)$ but marginalizing out $\bfz_i$, we have 
\begin{eqnarray}
\left( \bfx_i \mid s_i=l, k_i=k\right) \sim \mathrm{FG}_{\sigma}\left(\cdot \mid \bfc_l, r_l, \bfmu_k^l, \tau_k^l\right), \label{eq:fgmix2}
\end{eqnarray}
following the FG-kernel described in (\ref{FG}). 
Further marginalizing out $(s_i,k_i)$ leads to the simple two-layer FG kernel mixture 
\begin{eqnarray}
\bfx_i \sim \sum_{l=1}^{\infty} \lambda_l \sum_{k=1}^M \pi_{k}^l \mathrm{FG}_{\sigma}\left(\cdot \mid\bfc_l, r_l, \bfmu_k^l, \tau_k^l \right). \label{eq:fgmix3}
\end{eqnarray}
The proposed representation of $\bfx_i$ provides
a very flexible setup accommodating mixtures of multimodal spherical densities, and is suitable to approximate densities concentrated near an arbitrary curved support. 
A representative density of $x$, characterized as above, is shown in Figure \ref{fig1} (c).

A Bayesian model is completed with priors for the two levels of mixture weights and the kernel parameters, as described in the next subsection.

\subsection{Prior Specifications}
For the mixture distribution over spheres, we use a Dirichlet process prior (DP, \cite{Ferguson1973}, \cite{Antoniak}) with concentration parameter $\alpha$ and base measure $G_0=N\left({\bf 0},\sigma^2_c I\right)\times N_{+}\left(\mu_r,\sigma_r^2\right) $, where $\sigma_c^2,\mu_r, \sigma_r^2$ are positive constants and $N_+$ denotes a Gaussian density truncated to $(0,\infty).$  The DP prior induces a stick-breaking process  \citep{Sethuraman} on the probability weights $\boldsymbol{\lambda}=\left(\lambda_1,\lambda_2,\ldots \right)$.
For the $l^{th}$ sphere, the von-Mises kernel weights $\bfpi^l=\left(\pi^l_1,\ldots, \pi^l_M \right)$ follow a Dirichlet prior with parameters $a_0,\ldots,a_0$ for some fixed constant $a_0$.  For each sphere and each kernel, the parameters of the von-Mises Fisher density, $\left(\bfmu^{l}_m,\tau^{l}_m\right)$, are assigned the conjugate prior proposed in \cite{vmf_conjugate}, 
$$ p\left(\bfmu,\tau\right) \propto \left\{\frac{\tau^{d/2-1}}{I_{d/2-1}(\tau)} \right\}^a \exp\left(b\tau \bfmu_0^{\prime}\bfmu\right),$$
with hyper-parameters $\bfmu_0$, $a$ and $b$ for some $a>b>0$. Given $\tau$, the marginal density of $\bfmu$ follows a von-Mises Fisher density. However, the marginal density of $\tau$ does not have a standard form.  The variance of measurement error $\sigma^2$ is assigned an inverse-gamma prior with parameters $a_\sigma,b_\sigma >0$, where $a_\sigma$ and $b_\sigma$ are constants.
The choices of the hyperparameters are discussed in Section \ref{sec:4}.

\subsection{Posterior Computation}\label{sec:2}
Throughout this section we use the following conventions:
(i) The sphere and kernel indicators of the $i^{th}$ observation are denoted by $s_i$ and $k_i$, respectively, and the sphere and kernel indices are denoted by $l$ and $k$, $l=1,2,\ldots$, $k=1,\ldots, M$;
(ii) The number of observations in the $l^{th}$ sphere is denoted by $n^l$, and that in the $k^{th}$ kernel of the $l^{th}$ sphere is indicated by $n^l_k$; 
(iii) We let $\bfro_k^l=\left(\bfc_l,r_l,\bfmu_k^l,\tau_k^l\right)$ denote the parameters of the $k^{th}$ kernel on the $l^{th}$ sphere.

We develop a Metropolis within Gibbs sampler using the characterization of the density in (\ref{eqn:FGkernel}). The sampler is straightforward to implement, involving the following simple updating steps from standard distributions and no need for algorithm tuning. 
\paragraph{Step 1} {\it Updates on sphere labels.} We follow \citet[Algorithm 8]{Neal_2000} to update the sphere labels, $s_1,\ldots,s_n$.  Excluding the $i^{th}$ observation, suppose there are $L$ spheres represented in the data having parameters $\bfro_k^l$, $\pi_k^l$ and $\sigma^2$ for $l=1,\ldots,L$ and $k=1,\ldots,M$. Fix a number $J$, and create $J$ additional empty sphere labels, $L+1,\ldots,L+J$. If the $i^{th}$ observation was the only data point allocated to sphere $L^{\prime}$ in the last iteration, then attach the corresponding $\bfro_k^{L^\prime}$ and $\pi_k^{L^\prime}$ to the first empty sphere label, and associate the remaining empty sphere labels with values of $\bfro_k^l$ and $\pi_k^l$ generated randomly from the base distribution. We draw a new value of $s_i$ with 
\begin{equation}
P(s_i=l) \propto \begin{cases}
n_{-i}^l \sum_{k=1}^{M} \pi_k^l \mathrm{FG}_\sigma\left(\bfx_i\mid\bfro_k^l\right) & \mbox{for $l=1,\ldots,L$}, \\ 
\alpha J^{-1} \sum_{k=1}^{M} \pi_k^l \mathrm{FG}_\sigma\left(\bfx_i\mid\bfro_k^l\right) & \mbox{for $l=L+1,\ldots,L+J$},\label{eq_12}
\end{cases} 
\end{equation}
where $n_{-i}^l = \sum_{j \neq i} 1\left(s_j=l\right)$ and $\mathrm{FG}_\sigma\left(\bfx_i\mid\bfro\right)$ is as given in (\ref{FG}). 
\paragraph{Step 2} {\it Conditional posteriors of centers and radii.}
The conditional posterior of the center of the $l^{th}$ sphere $\bfc_l$ follows a Gaussian density with location parameter
$\bfmu_{c\mid\cdot}^l$ and scale parameter $\sigma^{l~2}_{c\mid\cdot} I$, where
$$ \bfmu_{c\mid\cdot}^l =\sigma_{c\mid\cdot} \times \left\{\frac{1}{\sigma^2} \sum_{i=1}^{n} I\left(s_i=l\right)\left(\bfx_i-r_l\bfy_i\right) \right\}~~\mbox{and}~~ \sigma^{l~2}_{c\mid\cdot}=\left(\frac{n^l}{\sigma^2}+\frac{1}{\sigma_c^2} \right)^{-1}. $$	
Similarly, the conditional posterior of the radius of the $l^{th}$ sphere $r_l$ follows a  positive Gaussian density with location parameter
$\mu_{r\mid\cdot}^l$ and scale parameter $\sigma^{l~2}_{r\mid\cdot} $, 
$$ \mu_{r\mid\cdot}^l =\sigma_{r|\cdot} \times \left\{\frac{\mu_r}{\sigma^2_r} +\frac{1}{\sigma^2} \sum_{i=1}^{n} I\left(s_i=l\right)\bfy_i^{\prime}\left(\bfx_i-\bfc_l\right)\right\}~~\mbox{and}~~ \sigma^{l~2}_{r\mid\cdot}=\left(\frac{n^l}{\sigma^2}+\frac{1}{\sigma_r^2} \right)^{-1}. $$
\paragraph{Step 3} {\it Updates on von-Mises Fisher kernel allocations and weights.} We first update the kernel allocation $k_i$, for $i=1,\ldots,n$, according to 
$$P(k_{i}=k\mid s_i=l,\bfy_i,\bfmu,\tau) =\frac{\pi_k^l~\mathrm{FG}_\sigma\left(\bfx_i\mid\bfc_l,r_l,\bfmu^l_k,\tau^l_k\right) }{\sum_{m=1}^{M} \pi_m^l~\mathrm{FG}_\sigma\left(\bfx_i\mid\bfc_l,r_l,\bfmu^l_m,\tau^l_m\right)}. $$
Then, the posterior distribution of $\left(\pi_1^l,\ldots,\pi_M^l\right)$ is Dirichlet with parameters $a_0+n^l_{1},\ldots,a_0+n^l_{M}$.
\paragraph{Step 4}{\it Update on spherical coordinates.} 
The latent spherical location, $\bfy_i$, corresponding to the $i^{th}$ observation $\bfx_i$, is updated from its conditional posterior density, which is a von-Mises Fisher density with parameters $\bfmu_{y\mid\cdot}$ and $\tau_{y\mid\cdot}$, where 
$$\bfmu_{y\mid\cdot}=\tau_{y\mid\cdot}^{-1} \left\{ \frac{r_{s_i}}{\sigma^2} \left(\bfx_i-\bfc_{s_i}\right)+\tau_{k_i}^{s_i} \bfmu_{k_i}^{s_i} \right\} \quad \mbox{and}\quad \tau_{y\mid\cdot}=\left\| \frac{r_{s_i}}{\sigma^2} \left(\bfx_i-\bfc_{s_i}\right)+\tau_{k_i}^{s_i} \bfmu_{k_i}^{s_i} \right\| . $$  
\paragraph{Step 5} {\it Update on $\sigma^2$.} The inverse-gamma prior of $\sigma^2$ is conjugate yielding an inverse-gamma posterior with parameters $a_{\sigma\mid\cdot}$ and $b_{\sigma\mid\cdot}$ where

{\centering
	$a_{\sigma\mid\cdot}= a_{\sigma}+ \displaystyle \frac{nd}{2} \quad \mbox{and}\quad b_{\sigma\mid\cdot}= b_{\sigma}+ \displaystyle \frac{1}{2} \displaystyle \sum_{i=1}^{n} \left(\bfx_i-\bfc_{s_i}-r_{s_i} \bfy_i\right)^{\prime} \left(\bfx_i-\bfc_{s_i}-r_{s_i} \bfy_i\right).$ \par}
\paragraph{Step 6} {\it Update on parameters of von-Mises Fisher distributions.} Finally, for each sphere, $l$, and each kernel, $k$, the vMF parameters are updated based on the observations associated with that sphere-kernel combination. The conditional prior on $\bfmu^l_k$ given $\tau^l_k$ is von-Mises Fisher, and is conjugate yielding a vMF posterior with parameters $\bfmu_{0\mid\cdot}$ and $\tau_{\mu\mid\cdot}$ where 
$$\bfmu_{0\mid\cdot}=\tau_{\mu\mid\cdot}^{-1} {\bf w}_k^l ~~ \mbox{and} ~ \tau_{\mu\mid\cdot}=\left\|{\bf w}^l_k \right\|, ~~ \mbox{with}~~ {\bf w}^l_k=\tau_k^l\left\{b\bfmu_0+\sum_{i=1}^{n} I\left(s_i=l,k_i=k\right)\bfy_i \right\}.$$ 
The conditional posterior of $\tau^l_k$ given other parameters is
$$ p_k^l\left(\tau\mid\cdot \right)\propto \left\{ \frac{\tau^{d/2-1}}{I_{d/2-1}(\tau)}\right\}^{a+n^l_k} \exp\left[\tau \left\{b\bfmu_0+\sum_i I\left(s_i=l,k_i=k\right) \bfy_i\right\}^{\prime}\bfmu_k^l \right]. $$
 We apply a Metropolis-Hastings algorithm with an independence sampler, considering a $\mathrm{Gamma}\left(2,u^l_k\right)$ proposal distribution. The details on choice of $u^l_k$ are given in Section \ref{sec:4}.

\section{Asymptotic Properties}\label{sec:3}
In this section, we show the Kullback-Leibler (KL) support property of the proposed prior, which implies weak posterior consistency (see \cite{schwartz1965}). 

Let $f_0$ be the true density function. Under the mixture of FG-kernel approach, the proposed density function $f$ is described in (\ref{eq:fgmix3}). For simplicity we show the KL property holds under a restrictive version of our model in which the same vMF parameters are re-used across the different spheres.  Under this assumption, we have 
\begin{eqnarray}
f_{P,\sigma}(\bfx)\coloneqq\int_\bfz \phi_{\sigma^2}(\bfx-\bfz)\left\{ \int \frac{1}{r^d} \sum_{j=1}^{M} \pi_j f_{\mathrm{vMF}}\left( \frac{\bfz-\bfc}{r} \,\middle\vert\,  \bfc,r ,\bfmu_j,\tau_j \right) dP \right\} d\bfz, \label{eq_1}
\end{eqnarray}
where the mixing distribution of $(\bfc,r,\bfpi)$, $P$, follows a DP prior with base measure $N\left(0,\sigma^2_c I_d\right)\times N_\delta\left(\mu_r,\sigma^2_r\right)\times \mathrm{Dir}\left(p,\ldots,p \right)$. Here $\{\bfmu_1, \ldots,\bfmu_M\}$ and $\{\tau_1,\ldots, \tau_M\}$ are prefixed quantities. The hyperparameters $\sigma_c^2$, $\mu_r$, $\delta$ and $\sigma_r^2$ are assumed to be elicited a priori.  

First, we state a lemma which makes the mixture density in (\ref{eq:fgmix3}) more theoretically tractable.  

\begin{lemma}\label{lm:3}
	For any  $\bfro=(\bfc,r,\bfmu,\tau)$ and $\sigma$, let the FG kernel be as in (\ref{FG}), then the following holds: 
	\begin{enumerate}[a.]
		\item For $d>2$, $\displaystyle(2\pi \sigma^2)^{-d/2} \exp\left\{-\displaystyle\frac{1}{2\sigma^2} \left(\|\bfx-\bfc\|+r \right)^2\right\} \leq \mathrm{FG}_\sigma\left(\bfx\mid \bfro\right)$
		
		       \hspace{3 in}$\leq 2 (2\pi \sigma^2)^{-d/2} \exp\left\{-\displaystyle\frac{1}{2\sigma^2} \left(\displaystyle\|\bfx-\bfc\| - r \right)^2\right\}.$
		\item For $d=2$, the above inequality holds with left and right hand sides multiplied by $\exp(-2\tau)$ and $\exp(2\tau)$, respectively. 
	\end{enumerate}
\end{lemma}

Let the true density be $f_0$, and $KL\left(f_0,f\right)=\int f_0 \log\left(f_0/f\right)$ be the KL divergence between $f_0$ and any density $f$. Let the KL neighborhood of $f_0$ of size $\varepsilon$ be $\mathcal{K}_{\varepsilon} (f_0)=\left\{f: KL\left(f_0,f\right)<\varepsilon \right\}$. We say that $f_0$ is in KL support of $\Pi$, written as $f_0\in KL\left(\Pi\right)$, if $\Pi\left\{ \mathcal{K}_{\varepsilon}\left(f_0\right) \right\}>0$, for every $\varepsilon>0$.

Next we consider the assumptions under which our results hold.
\begin{enumerate}[A.]
	\item The true density $f_0(\bfx)<M_0$ for some constant $M_0>0$ for all $\bfx\in \mathbb{R}^d$.
	\item The true density $f_0$ satisfies: $ |f_0(\bfx+\bfy)-f_0(\bfx)|\leq L(\bfx) \exp\left(s\|\bfy\|^2 \right) \|\bfy\|^{r_0}$,
	for any $\bfx,\bfy\in \mathbb{R}^d$, some $s,r_0 \geq 0$, and a non-negative function $L$ in $\mathbb{R}^d$ such that $L(\bfx) < M^{\prime} f_0(\bfx)$, for some large constant $M^{\prime}$.
	\item The $q^{th}$ order moments of $\bfx$ are finite, for some $q>2$.
	\item There exists a fixed point $\bfx_0$ and some fixed $\delta>0$ such that $\inf_{\|\bfx-\bfx_0\|<\delta}f_0(\bfx)=\phi_\delta>0$. 
\end{enumerate}

Assumption A is common in the literature (see, for e.g., \cite{WG2008}). Assumption B is weaker than the usual H\"older continuity with $\alpha=r_0$. For $r_0<1$, $f_0$ satisfying assumption B belongs to a locally $r_0$-H\"older class with envelope $L$ (see \cite{STG2013}).  Assumptions C and D are weak conditions. 

\begin{theorem}\label{thm:1}
	Suppose the true density $f_0$ satisfies assumptions A-D, and the prior $\Pi$ is as given in (\ref{eq_1}). Then $\Pi\left\{\mathcal{K}_{\varepsilon}(f_0) \right\}>0$ for any $\varepsilon>0$.
\end{theorem}

\begin{remark}
	In Theorem \ref{thm:1}, we consider the mixing distribution of $(\bfc,r,\bfpi)$ is $P$, $P\sim DP$, and $(\bfmu,\tau)$ varies on a fixed set.  
	One can trivially extend Theorem $\ref{thm:1}$ to the case where the mixing distribution of $(\bfc, r, \bfpi,\bfmu_1,\cdots,\bfmu_M,\tau_1,\cdots, \tau_M)$ is $P$, if $\tau$ is restricted  to a compact set $[0,\tau_{\max}]$.
\end{remark}

\section{ Simulation Experiments}\label{sec:4}
In this section we carry out a variety of simulation experiments to validate our method. Our focus is on datasets which are concentrated near a manifold. We compare our method with frequentist kernel density estimation (KDE) and Dirichlet process mixtures of Gaussians (DPMG).

{\it Our method:}    The algorithm is described in Section \ref{sec:2}. We keep the choices of precision parameter $\alpha$, and number of empty spheres $J$ fixed, and consider two choices of $\alpha/J$, $\{1,2\}$. 
The number of von-Mises Fisher (vMF) kernels within each sphere is set to $5,10$ or $15$ based on the complexity of the data. The parameter $a_0$ in the prior for the vMF kernel weights $\boldsymbol{\pi}$ is set to $1$. The hyperparameters of the inverse-gamma prior of $\sigma^2$ are set to $\mathrm{rate}=1$ and $\mathrm{scale}=0.01$ to make the prior weakly informative. The hyperparameters of the centers and radii are set to $\sigma_c^2=1$, $\mu_r=1$ and $\sigma_r=5$.
The $\bfmu_j$s are given a vMF prior with parameters $\bfmu_0=\left(1/\sqrt{d},\ldots, 1/\sqrt{d}\right)$ and $a=1$ with $b=0.1$. A small value of $b$ makes the prior less informative. For the independence sampler applied to update $\tau^l_m$, we use a gamma proposal distribution with shape $2$ and rate $2/\hat{\tau}^l_m$. Here $\hat{\tau}^l_m$ is an approximation to the maximum likelihood (ML) estimate of $\tau^l_m$, $\hat{\tau}^l_m=\bar{t} \left(d-\bar{t}\right)/\left(1-\bar{t}^2\right)$, where $\bar{t}=\| \sum_j \bfy^l_{j,m}  \|/n^l_m$ (see \cite{sra2012} for details on ML estimation of vMF parameters). Here $\bfy_{j,m}^l$ denotes the latent spherical coordinates of the $j^{th}$ observation belonging to the $l^{th}$ sphere and $m^{th}$ kernel. This proposal provides acceptance rates between $20\%$-$40\%$ in our experiments.

{\it Other methods:} For comparison we implement DMPG using the `DirichletProcessMvnormal' function in the `dirichletprocess' R-package with all default settings. Multivariate Gaussian kernels are used, with the base measure of the DP corresponding to a multivariate normal-inverse Wishart distribution and with the 
concentration parameter $\alpha$ given a gamma prior.  Algorithm 8 in \cite{Neal_2000} is used for sampling.  
For kernel density estimation (KDE) we use the `kde' function with default settings in the `ks' R-package.  This corresponds to using multivariate Gaussian kernels with covariance/bandwidth matrix $H$.  By default `ks' selects the bandwidth $H$ by direct plug-in methodology, as described in  \cite{plug_in_bandwidth1} and \cite{plug_in_bandwidth2} for univariate and multivariate cases, respectively.  
Most of the packages including `ks' do not allow data of dimension greater than 6 (see, for e.g., \cite{kde_pack_info}). Thus for cases with dimension greater than 6, we use the `kdevine' package. This package estimates marginal densities using KDE, and then uses the VINE copula to obtain the multivariate joint density estimate.   

We split the validation into two categories: density estimation and density-based classification. Below we briefly describe the measures used to assess performance. 

\vskip5pt
\noindent (i) {\it Histogram estimate.} We generate $n$ samples, say $\bfz_1,\ldots,\bfz_n$, from the predictive distribution, $n$ being the training size; and choose $N$ points, $\bfx_1,\ldots, \bfx_N$, randomly with replacement from the training data. Fix a neighborhood size $\delta$. Let $W_{j,\delta}^{z}=\sum_{i=1}^{n}I(\|\bfz_i-\bfx_{j}\|<\delta)$ be the number of predictive points and $W_{j,\delta}^{x}=\sum_{i=1}^{n}I(\|\bfx_i-\bfx_{j}\|<\delta)$ be the number of training points in the $\delta$-neighborhood of $\bfx_j$, $j=1,\ldots,N$. Then the histogram estimate is the mean absolute difference $\mathrm{MAD}_\delta=\sum_{j=1}^{N}| W_{j,\delta}^{z} - W_{j,\delta}^{x}|/N$.

\vskip5pt
\noindent (ii) {\it Likelihood estimate:} Splitting the data into training and test sets, we estimate the density $f$ using the training set. We find the predictive likelihood of each test point $\bfx$, $\hat{f}(\bfx)$. Out-of-sample goodness of fit is summarized via the estimated log-likelihood, $\sum_{j}\log \hat{f}(\bfx_j)$, or boxplots of $\hat{f}(\bfx_j)$.

\vskip5pt
\noindent (iii) {\it Classification Accuracy.} Using the training data, we separately estimate the density of the features within each class. Letting $\hat{f}_j(\bfx)$ denote the density within class $j$, we assign the test data $\bfx^{*}$ to the class having the highest $\hat{f}_j\left(\bfx^{*} \right)$. We then report the classification accuracy as the proportion of correct classifications in the test data. 

Below we consider three density estimation examples along with a classification example. 

\begin{center}
	\begin{table*}[!h]
		\renewcommand{\arraystretch}{1.2}
		\begin{minipage}[c]{6.5 cm}
			{\footnotesize
				\caption{\emph{MAD$_\delta$} for different choices of $\delta$ for 3 competing methods in Euler Spiral dataset}
				\label{tab1}
				\begin{tabular}{|p{.9 cm}|p{1.1 cm} p{1.1 cm} p{1.2 cm} |}
					\hline 
					$\delta$ &   KDE  & DPMG  &   FG-mixture  \\ \hline 
					0.005 & 0.928 & 0.964 & {\bf 0.924} \\
					0.010 & 2.816 & 3.242 & {\bf 1.924} \\
					0.015 & 4.512 & 5.624 & {\bf 2.428} \\
					0.020 & 5.836 & 7.628 & {\bf 2.930} \\
					0.025 & 6.886 & 9.906 & {\bf 3.646} \\
					0.030 & 7.714 & 11.524 & {\bf 4.118} \\
					0.035 & 8.202 & 13.084 & {\bf 4.660} \\ \hline 
			\end{tabular} }
			
		\end{minipage}\qquad\qquad
		\begin{minipage}[c]{6.5 cm}
			\renewcommand{\arraystretch}{1.2}
			{\footnotesize
				\caption{\emph{MAD$_\delta$} for different choices of $\delta$ for 3 competing methods in `Olympic rings' dataset}
				\label{Orings}
				\begin{tabular}{|p{.9 cm}|p{1.1 cm} p{1.1 cm} p{1.2 cm} |}
					\hline 
					$\delta$ &  KDE  &  DPMG  &  FG-mixture  \\ \hline
					0.025 	&	2.102	&	2.242 	&	{\bf 1.646} \\
					0.050	&	4.256 	&	4.782	&	{\bf 2.688}	\\
					0.075	&	6.014	&	6.858	&	{\bf 3.526} \\
					0.100	&	6.838	&	8.392	&	{\bf 4.118} \\
					0.125	&	7.316 	&	9.906	&	{\bf 4.738} \\
					0.150	&	8.072	&	11.086	&	{\bf 5.350} \\
					0.175	&	8.850	&	13.122	&	{\bf 5.278} \\ \hline 
			\end{tabular} }
		\end{minipage}
	\end{table*}
\end{center}

\begin{figure}\label{ESp_scatter}
	\begin{center}
		\includegraphics[height=1 in, width=5 in]{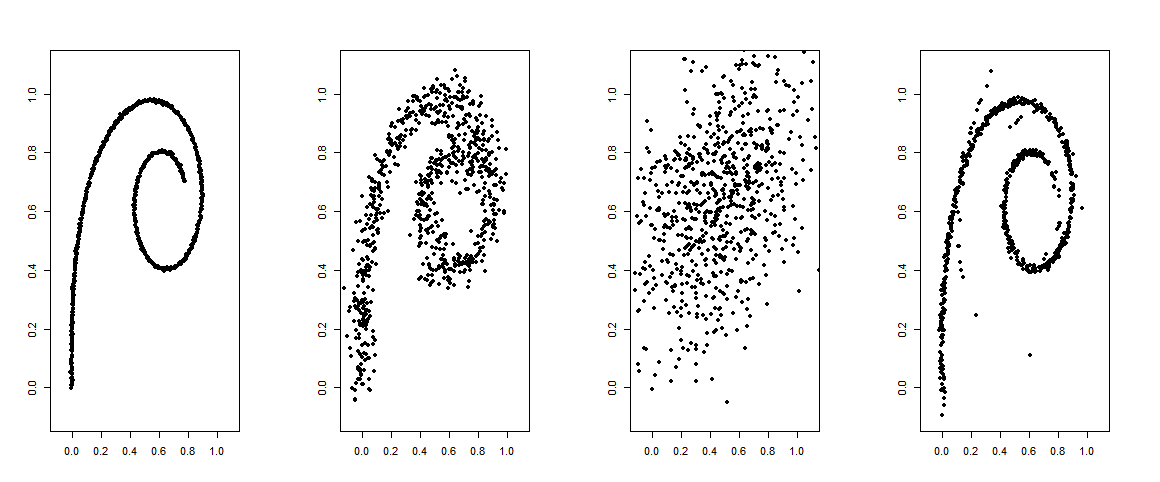}
		\vspace{-1cm}
		\includegraphics[height=1.5 in, width=5 in]{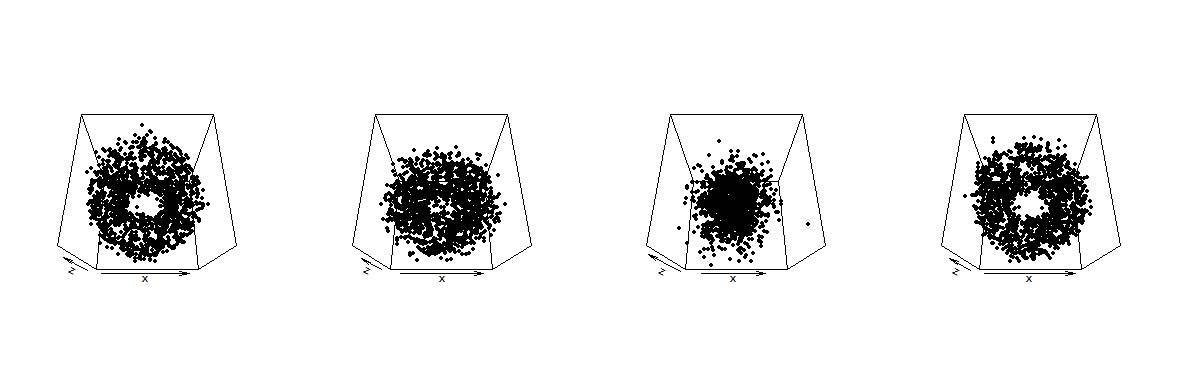}
		\caption{Row-wise: (i) Scatter plot of the Euler spiral dataset (first column) and $500$ sample points from predictive densities of KDE (second column), DPMG (third column) and FG-mixture (last column). (ii) Scatter plot of the Torus dataset (first column) and $1500$ sample points from predictive density of KDE (second column), DPMG (third column) and FG-mixture (last column).}\label{fig3}
	\end{center}
\end{figure}

\begin{figure}
	\begin{center}
		\includegraphics[height=1.8 in, width=1.9 in]{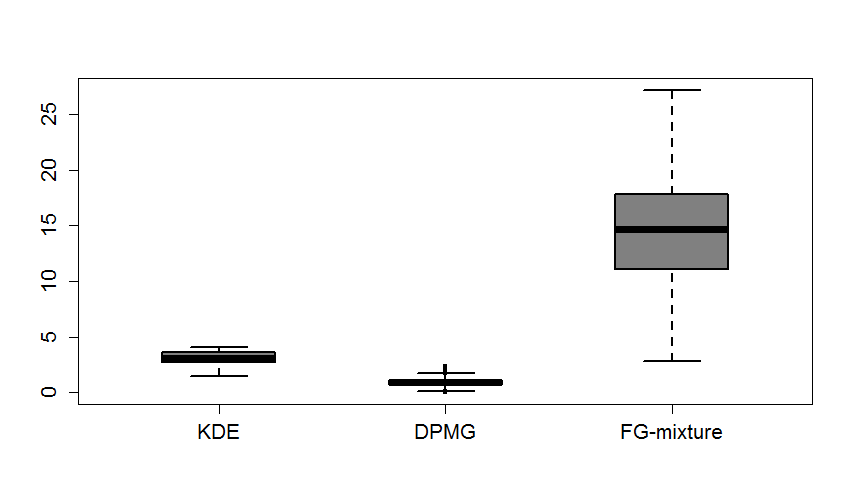}
		\includegraphics[height=1.8 in, width=1.9 in]{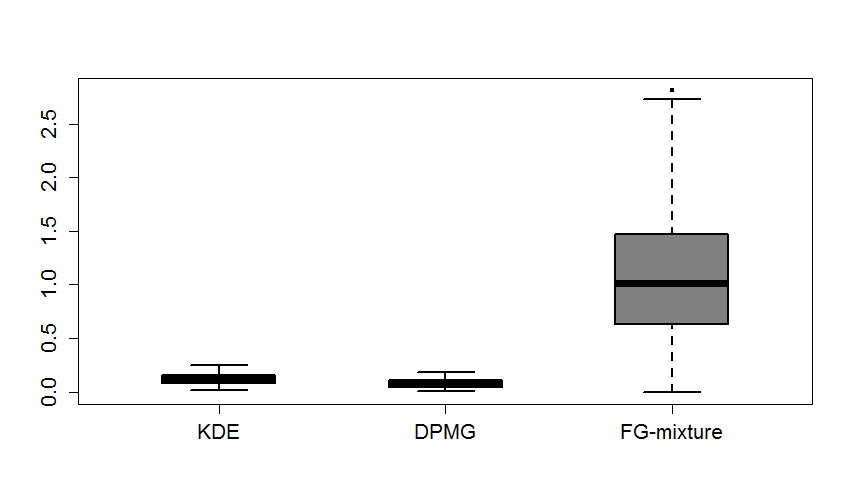}
		\includegraphics[height=1.8 in, width=1.9 in]{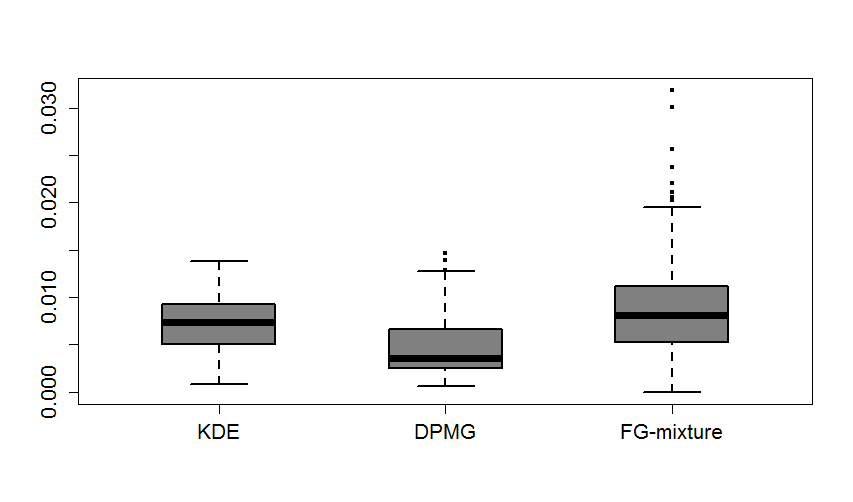}
		\caption{Box-plot of likelihood of test samples for three competing methods for Euler spiral data (left), Olympic rings data (middle) and Torus dataset (right panel) }\label{fig4}
	\end{center}
\end{figure}

\paragraph{Euler Spiral} An Euler spiral is a curve whose curvature changes linearly with its curve length. We consider a sample of size $500$ from the Euler spiral, and add a Gaussian noise of variance $0.001^2 I$. The mean absolute differences \emph{MAD$_\delta$} for $N=2000$ and different values of $\delta$ are given in Table \ref{tab1}. Box-plots of estimated likelihoods of $300$ test points are given in Figure \ref{fig3}. A Scatter plot of an Euler spiral data and that of $500$ predictive points from each method are shown in Figure \ref{fig4}.    

\paragraph{Olympic Rings} We generate $1500$ datapoints from 5 circles forming Olympic rings. The five circles have centers at $(0,0)$, $(2.125,0)$, $(4.25,0)$, $(1.125,-1)$ and $(3.25,-1)$, respectively, and radii $1$. We generate $i\times 100$ points from the $i^{th}$ circle, and add a Gaussian noise component with sd $0.01$ for each component. The values of \emph{MAD$_\delta$} for $N=2000$ and different values of $\delta$ are given in Table \ref{Orings}, and box-plots of estimated likelihoods are shown in Figure \ref{fig4}.	

\paragraph{Torus} A torus (or {\it torus of revolution}) is a surface of revolution generated by revolving a circle in three-dimensional space about an axis coplanar with the circle. The axis of revolution does not touch the circle. We generate a dataset of $n=1500$ points on a torus using the `torus' function in the `geozoo' R-package. The radius of the larger circle is set to $3$, that of the smaller circle is set to $1$, and a Gaussian noise with variance $0.1$ is added. The values of \emph{MAD$_\delta$} for $N=1000$ and different values of $\delta$ are given in Table \ref{tab3}, figures of the Torus dataset and $1500$ predictive points are shown in Figure \ref{fig3}, and box-plots of estimated likelihoods are in Figure \ref{fig4}.

\begin{figure}[h]
	\begin{center}
		\includegraphics[	height=1.5 in, width=5.5 in]{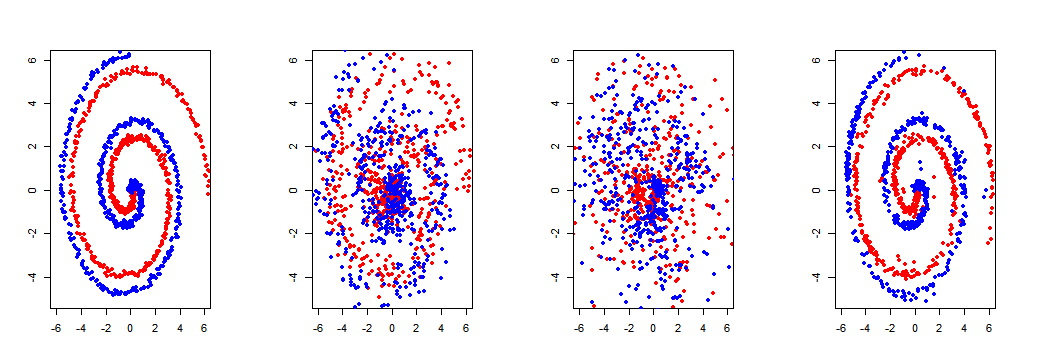}
		\caption{Scatter plot of the `two spirals' dataset (first column) and $750$ predictive sample points from each spiral of KDE (second column), DPMG (third column) and FG-mixture (last column). }\label{two_spiral}
	\end{center}
\end{figure}

\paragraph{Two-Spiral} The last example is that of two spirals with varying curvature. The spirals intersect only at the center at the point of maximum curvature. As the curves move outward, the curvature of the spirals decreases simultaneously.
The classification accuracies for different training test splits are given in Table \ref{tab2}, and the scatter plot of the actual dataset and $750$ predictive points from each spiral are shown in Figure \ref{two_spiral}. 

{\centering
	\begin{table*}[h]
		\renewcommand{\arraystretch}{1.2}
		\begin{minipage}[c]{6.2 cm}
			{\footnotesize
				\caption{The \emph{MAD$_\delta$} for different choices of $\delta$ for the `Torus' dataset}
				\label{tab3}
				\begin{tabular}{|p{0.8 cm}|p{1 cm} p{1 cm} p{1.1 cm} |}
					\hline
					$\delta$ & KDE  & DPMG  &  FG-mixture  \\ \hline
					0.1	&	0.121 & {\bf 0.096} & 0.117 \\
					0.2	&	0.626 & {\bf 0.552} & 0.625 \\
					0.3	&	1.255 & 1.381 	&	{\bf 1.204}\\
					0.4	&	2.127 & 2.653 	&	{\bf 1.982}\\
					0.5	& 	2.829 & 3.991 & {\bf 2.611} \\
					0.6	&	 3.703 & 5.754 & {\bf 3.121} \\
					0.7 &     5.515 & 11.071 & {\bf 5.340} \\ \hline 
			\end{tabular} }
		\end{minipage}  \qquad 
		\begin{minipage}[c]{6.8 cm}
			{\footnotesize
				\caption{The accuracy for different training test split of the two-spiral dataset}
				\label{tab2}
				\begin{tabular}{|p{1 cm}|p{1 cm} |p{.75 cm} p{.75 cm} p{1.1 cm} |} 
					\hline 
					Training & Test & \multicolumn{3}{|c|}{Classification accuracy} \\ 
					size & size &  KDE  &  DPMG  &  FG-mixture \\ \hline
					50  & 100 &0.655 & 0.550 & {\bf 0.945} \\ 
					100 & 100 & 0.695	& 0.610	&	{\bf 0.980} \\ 
					150 & 100 &	0.835 &	0.580 &	{\bf0.980}	\\ 
					200 & 100 &	0.940&	0.645&	{\bf 0.985}	\\ 
					250 & 100 &	0.975& 	0.695 &	{\bf 0.980}	\\ 
					300 & 100 &	0.955 & 0.655 & {\bf 0.990}	\\ \hline 
			\end{tabular} }
		\end{minipage}\qquad\qquad
	\end{table*} 
	\par}

The simulation results show almost uniformly better performance for the FG-mixture approach, with gains very dramatic in some cases.
\section{Applications}\label{sec:5}
In this section, we consider three datasets, {\it Galaxy}, {\it Gesture Phase} and {\it Balance scale}, for classification, and two datasets, {\it User knowledge modeling} and {\it Balance scale}, for density estimation. The results are given below. 

\begin{figure}[h]
	\begin{center}
		\includegraphics[height=5.5 cm, width=7.5 cm,trim={1cm 0 0 0},clip]{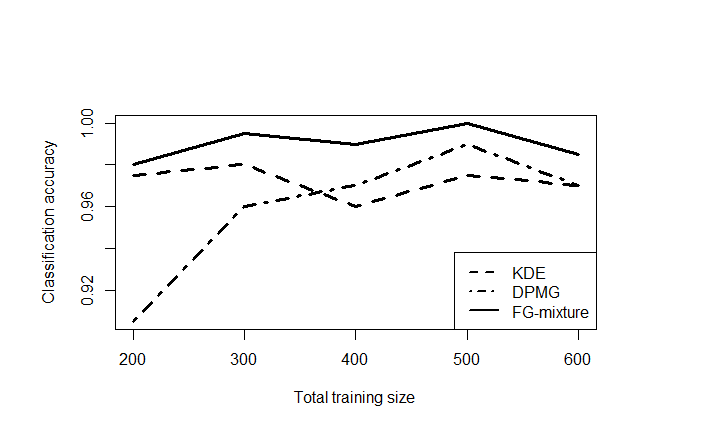}
		\includegraphics[height=5.5 cm, width=7.5 cm,trim={1cm 0 0 0},clip]{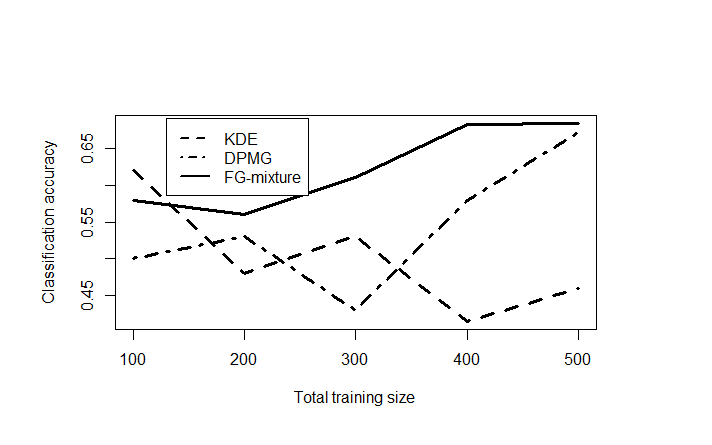}
		\caption{Classification accuracies of different methods for the Galaxy (left panel) and Balance scale data (right panel). The training and test sizes for Galaxy data are $(100,100)$, $(150,100)$, $(200,100)$, $(250,100)$ and $(300,100)$; and those for Balance scale data are $(50,100)$, $(100,100)$, $(150,100)$, $(200,88)$ and $(250,38)$ from each population. }\label{fig5}
	\end{center}
\end{figure}

\paragraph{Galaxy dataset} is available at \url{https://data.galaxyzoo.org/}. In the original Galaxy Zoo project (see \cite{GalaxyZoo} for details), volunteers classified images of Sloan Digital Sky Survey galaxies as belonging to one of six categories.
We consider the first dataset (Table 2), where the galaxies are classified as spiral, elliptical or merger (i.e., uncertain). 
This dataset has information on 10,000 galaxies, among which 2888 are spiral and 931 are elliptical. Among these samples we consider $100$ spiral and $100$ elliptical galaxies at random as the test set, and vary the size of the training set. We consider 5 predictors \emph{$\mathrm{P_{EL}}$}, \emph{$\mathrm{P_{CW}}$}, \emph{$\mathrm{P_{EDGE}}$}, \emph{$\mathrm{P_{DK}}$}, \emph{$\mathrm{P_{MG}}$} for classification. The classification accuracy is given in Figure \ref{fig5}, showing that it is improved significantly using FG mixtures.  

\paragraph{Gesture Phase dataset} is available at the UCI machine learning repository. It is composed of features extracted from 7 videos with people gesticulating, aimed at studying gesture phases. We use the processed table {\it a1\_va3}, which has $32$ attributes, and $5$ labels, viz., D, S, H, P, R. Among these $5$ labels we discard H as it has fewer instances, and compare each pair of labels D, S, P and R. The results are shown in Table \ref{tab5}. 

\begin{table*}[!h]
	\renewcommand{\arraystretch}{1.2}
	\begin{center}
		{\footnotesize
			\begin{threeparttable}
				\caption{Classification accuracies of the competing methods for the Gesture Phase data }
				\label{tab5}
				\begin{tabular}{|p{1 cm} p{1 cm}| p{0.8 cm} p{0.8 cm} p{1.2 cm}  |p{1.1 cm} p{1.1 cm}|  p{0.8 cm} p{0.8 cm} p{1.2 cm} |}
					\hline 
					\multicolumn{2}{|c|}{{\bf D against P}} & \multicolumn{3}{|c|}{ Classification Accuracy }  & \multicolumn{2}{|c|}{{\bf D against S}} & \multicolumn{3}{|c|}{ Classification Accuracy }  \\
					Train size \tnote{*}  &   Test size \tnote{*} & KDE &  DPMG  &   FG-mixture  & Train size * &  Test size * &  KDE &   DPMG  & FG-mixture  \\ \hline 
					50 & 100 &  ** &  0.490 & {\bf 0.585}  & 100 & 100 & 0.475  & 0.690 & {\bf 0.700}  \\
					100 & 63 & 0.508 &  0.603 & {\bf 0.706} & 200 & 100 &  0.495 & {\bf 0.740} & 0.700  \\ \hline
					\multicolumn{2}{|c|}{{\bf D against R}} & \multicolumn{3}{|c|}{ Classification Accuracy }   & \multicolumn{2}{|c|}{{\bf P against S}} & \multicolumn{3}{|c|}{Classification Accuracy }  \\ \hline
					50 & 100 & ** & 0.455 & {\bf 0.655} & 50 & 100 & ** &  0.505 &  {\bf 0.590}  \\ 
					100 & 91 & 0.500 & 0.725 &  {\bf 0.736}  &  100 & 63 & 0.500 & 0.548 & {\bf 0.579}  \\  \hline
					\multicolumn{2}{|c|}{{\bf P against R}} & \multicolumn{3}{|c|}{ Classification Accuracy }  & \multicolumn{2}{|c|}{{\bf S against R}} & \multicolumn{3}{|c|}{ Classification Accuracy }  \\ \hline 
					50 & 100 & ** & 0.530 & {\bf 0.585} & 50 & 100 & 0.495 & 0.475 & {\bf 0.550}  \\
					100 & 63 & 0.516 & 0.627 & {\bf 0.635} & 100 & 91 & 0.532 & 0.595 & {\bf 0.611} \\ \hline
				\end{tabular}
				\begin{tablenotes}
					\item [*] {\it Train size} and {\it Test size} refer to the training and test sizes of each population.
					\item [**] KDE estimates of density for all the test samples are zero under both the populations. 
				\end{tablenotes} 
		\end{threeparttable}}
	\end{center}
\end{table*}

\paragraph{Balance scale dataset}is generated to model psychological experimental results by \cite{balance_scale}, and is obtained from the UCI repository. Each example is classified as having the balance scale tip to the right, left, or balanced. The predictors are left-weight, left-balance, right-weight and right-balance. There are 288 instances of each of left and right tipped balance scales, and we consider classification between left and right tipped balance scales. The classification accuracies are given in Figure \ref{fig5}.

We further use Balance scale data for density estimation. Towards that we consider training samples of various sizes from the unlabeled population, and a test data of size 100. The sum of estimated log-likelihoods of the test data is provided in Table \ref{tab7}.   

\begin{center}
	\begin{table*}[!h]
		\begin{minipage}[c]{7.5 cm}
			\renewcommand{\arraystretch}{1.2}
			{\footnotesize
				\caption{Sum of log-likelihoods of test samples for competing methods in Balance scale data}
				\label{tab7}
				\begin{center}
					\begin{tabular}{|p{1 cm} p{1 cm}|p{1.2 cm} p{1.2 cm} p{1.2 cm} |}
						\hline 
						Training size & Test size &   KDE  &  DPMG  &  FG-mixture  \\ \hline
						50 	& 100 &	-645.20 & -611.72 & {\bf -574.93} \\
						100	& 100 &	-625.78 & -571.38 & {\bf -549.56}	\\
						150	& 100 &	-621.60 & -574.87 & {\bf -550.17} \\
						200	& 100 &	-605.33 & -567.37 & {\bf -551.23}  \\\hline 
					\end{tabular}
			\end{center}}
		\end{minipage} \hspace{.1 in}
		\begin{minipage}[c]{8 cm}
			\renewcommand{\arraystretch}{1.2}
			{\footnotesize
				\caption{Sum of log-likelihoods of test samples for competing methods in User knowledge modeling data}
				\label{tab6} \vspace{-.18 in}
				\begin{center}
					\begin{tabular}{|p{1 cm} p{1 cm}|p{1.1 cm} p{1.1 cm} p{1.2 cm} |}
						\hline 
						Training size  & Test size &   KDE  &  DPMG  &  FG-mixture  \\ \hline
						50  & 100 & -230.90 &  -27.94 & {\bf -25.69} \\
						100	& 100 & -805.63 & -713.39 & {\bf -712.81}	\\
						150	& 100 & -48.30  & 7.36  & {\bf 37.21} \\
						200	& 58 &  11.88 &  7.46 & {\bf 14.30} \\\hline 
					\end{tabular}
			\end{center}}
		\end{minipage}\qquad\qquad
	\end{table*} 
\end{center}

\paragraph{User knowledge modeling dataset} 
is about the students' knowledge status on the subject of Electrical DC Machines, and the data are obtained from the UCI repository. There are 5 attributes measuring the study time, repetition of different aspects and knowledge level, and 258 instances. The sum of estimated log likelihoods of test samples for varying training-test splits are given in Table \ref{tab6}

From these applications it is evident that FG mixtures clearly improve performance in both density estimation and classification, particularly in smaller samples. Also some of the improvements are quite dramatic.

\section{Discussion}\label{sec:6}

The main contribution of this article has been to introduce a simple generalization of the Gaussian distribution to allow the density to be concentrated near a curved support.  The resulting Fisher-Gaussian distribution has a simple analytic form, and we have demonstrated that it can lead to dramatically better performance in density estimation when used in place of usual Gaussian kernels.  This is particularly true when the data are concentrated near a non-linear subspace or manifold and when there are non-linear relationships among the variables being modeled, as is very commonly the case in applications.  There have been many different multivariate distributions proposed in the literature, motivated largely by limitations of the multivariate Gaussian distribution in terms of symmetry and light tails, but essentially no consideration of the curved support problem.  Multivariate distributions that can characterize data concentrated near a curved support tend to have a very complex form that is not analytically tractable, and hence there are considerable computational and interpretability hurdles in their implementation.  

There are a number of interesting directions for future research. Natural extensions include generalizing the Fisher-Gaussian distribution to accommodate a more flexible covariance structure.  One possibility is to include an arbitrary covariance in the Gaussian residual density instead of focusing on the spherical covariance case.  Another is to use a more flexible generalization of the von Mises-Fisher density on the sphere; several such generalizations are available in the literature (\cite{bingham},\cite{fisher_binghum}).  We have made initial attempts along these directions and have thus far been unable to obtain an analytically tractable form for the density of the data $\bfx_i$ marginalizing out the coordinates on the sphere $\bfy_i$.  It is possible to conduct computation without such a closed form, but it will be unwieldy.  

Another direction is to start with a density on a more flexible non-linear manifold than the sphere before adding Gaussian noise.  Possibilities include an ellipse, quadratic surface or hyperbolic space.  Unfortunately, it is not so straightforward to define tractable densities on such spaces, and it seems daunting to maintain analytic tractability in doing so.  One promising direction is to start with a Gaussian on a tangent plane to a manifold and then apply the exponential map to induce a density on the target manifold.  For a number of manifolds, it becomes possible to define the Jacobian and obtain a tractable form for the density on the manifold.  For certain cases, it may also be possible to add Gaussian noise and marginalize out the coordinates on the manifold, as we did for the FG distribution.  Otherwise, one can always rely on data augmentation algorithms for computation. 

As a final fascinating direction, it will be interesting to develop theory directly justifying the use of `curved support' kernels in density estimation.  We have focused on showing weak posterior consistency but it would be appealing to show that a faster rate of posterior concentration can be obtained by using FG kernels instead of spherical Gaussian kernels, at least when the true density is concentrated near a non-linear manifold with sufficiently large curvature.

\vspace{5pt}
{\noindent {\bf Acknowledgements} }\\
This research was partially supported by grant N00014-14-1-0245/N00014-16-1-2147 of the United States Office of Naval Research (ONR) and 5R01ES027498-02 of the United States National Institutes of Health (NIH).

\section{Appendix}\label{sec:7}
\subsection*{Proof of Lemma \ref{lm:1}}
\begin{proof}[ Proof of part (a).]
	Observe that 
	\begin{eqnarray*}
		&& \frac{1}{r^d} \int_{\RR^d}  \phi_{\sigma^2}(\bfx-\bfz) f_{\mathrm{vMF}}\left( \frac{\bfz-\bfc}{r} \,\middle\vert\,  \bfc,r ,\bfmu ,\tau \right) d\bfz  \\
		&&= \frac{1}{r^d} C_d(\tau) \left(2\pi\sigma^2\right)^{-d/2}\int_{\|\bfz-\bfc\|=r} \exp\left\{ -\frac{1}{2\sigma^2}\| \bfx-\bfz\|^2+ \tau \left( \frac{\bfz-\bfc}{r}\right)^{\prime}\bfmu \right\} d\bfz \\
		&&= \frac{1}{r^d}  \left[ C_d(\tau) \left(2\pi\sigma^2\right)^{-d/2}\int_{\|\bfz-\bfc\|=r} \exp\left\{ -\frac{1}{2\sigma^2}\left(\| \bfx-\bfc\|^2+r^2\right)\right. \right. \\
		&&\hspace{3.25 in} \left. \left. + \left(\frac{\bfz-\bfc}{r}\right)^{\prime}\left(\tau \bfmu +\frac{r(\bfx-\bfc)}{2\sigma^2} \right) \right\} d\bfz \right] \\
		&&=  C_d(\tau) \left(2\pi\sigma^2\right)^{-d/2}\int_{\|\bfw\|=1} \exp\left\{ -\frac{1}{\sigma^2}\left(\| \bfx-\bfc\|^2+r^2\right)   + \bfw^{\prime}\left(\tau \bfmu +\frac{r(\bfx-\bfc)}{2\sigma^2} \right) \right\} d\bfw \\
		&& = \frac{C_d(\tau )\left(2\pi\sigma^2\right)^{-(d/2)}}{C_d\left(\|\tau \mu +r(\bfx-\bfc)/\sigma^2\|\right)} 
		\exp\left\{ -\frac{1}{2\sigma^2} \left(\|\bfx-\bfc\|^2+r^2\right)\right\}.  \hspace{2 in } 
	\end{eqnarray*} 
	\noindent {\it \bf Proof of part (b).}
	Given $\bfro=(\bfc,r,\bfmu,\tau)$ we can write 
	$$\int_{\RR^d} \mathrm{FG}_{\sigma}(\bfx|\bfro) d\bfx= \int_{\RR^d} \int_{\RR^d} \frac{1}{r^d} \phi_\sigma(\bfx-\bfz) f_{\mathrm{vMF}}\left( \frac{\bfz-\bfc}{r} \,\middle\vert\,  \bfmu,\tau \right) d\bfz d\bfx.   $$
	Next we interchange the integrals of $\bfx$ and $\bfz$ as the integrand is non-negative and the integral is finite. After the changing the integral we get:
	$$\int_{\RR^d} \frac{1}{r^d} f_{\mathrm{vMF}}\left( \frac{\bfz-\bfc}{r} \,\middle\vert\,  \bfmu,\tau \right) \left\{ \int_{\RR^d} \phi_\sigma(\bfx-\bfz) d\bfx \right\} d\bfz   = \int_{\RR^d} \frac{1}{r^d} f_{\mathrm{vMF}}\left( \frac{\bfz-\bfc}{r} \,\middle\vert\, \bfmu,\tau \right) d\bfz=1.$$
\end{proof}
\subsection*{Proof of Lemma \ref{lm:3}}
\begin{proof}
	Let $r (\bfx-\bfc)/\sigma^2=\bfw$.  Note that $$\frac{C_d(\tau)}{C_d \left(\left\|\tau \bfmu +\bfw \right\| \right)}
	= \frac{\|\tau \bfmu +\bfw\|^{-(d/2-1)} I_{d/2-1}(\tau \bfmu +\bfw)}{(\tau )^{-(d/2-1)} I_{d/2-1}(\tau)}.$$ 
	By \cite[Equation (2.2)]{MBFbound}, for $d> 2$,	$$\exp \left\{-\left|\|\tau \bfmu +\bfw\|-\tau \right| \right\}
	\leq  \frac{(\tau \bfmu + \bfw)^{-(d/2-1)} I_{d/2-1}(\tau \bfmu +\bfw)}{(\tau)^{-(d/2-1)} I_{d/2-1}(\tau)}  \leq \exp \left(\left|\|\tau \bfmu +\bfw\|-\tau \right| \right).$$
	Next, observe that
	\begin{eqnarray*}
		\left|\tau -\|\bfw\|\right|-\tau &\leq \left| \|\tau \bfmu +\bfw\|-\tau  \right| &\leq  \|\bfw\| \qquad \mbox{when}\quad \|\tau \bfmu +\bfw\|>\tau, \\
		-\|\bfw\| &\leq \left| \|\tau \bfmu +\bfw\|-\tau  \right| &\leq  \tau - \left|\tau - \|\bfw\|\right| \qquad \mbox{when}\quad \|\tau \bfmu +\bfw\|\leq \tau,
	\end{eqnarray*}
	as $\|\bfmu \|=1$ and $|\|{\bf a}\|-\|{\bf b}\||\leq \|{\bf a}+{\bf b}\|\leq \|{\bf a}\|+\|{\bf b}\|$. This reduces to
	\begin{eqnarray*}
		-\|\bfw\| &\leq \left| \|\tau \bfmu+\bfw\|-\tau  \right| &\leq  \|\bfw\| \qquad \mbox{when}\quad \|\bfw\|\leq \tau<\|\tau \bfmu +\bfw\| \\
		&& \qquad \qquad\mbox{or}\quad \tau \geq \max\left\{\|\bfw\|,\|\tau \bfmu +\bfw\|\right\}, \\
		\|\bfw\|-2\tau  &\leq \left| \|\tau \bfmu +\bfw\|-\tau   \right| &\leq  \|\bfw\| \qquad \mbox{when}\quad \tau <\min\left\{\|\bfw\|,\|\tau \bfmu +\bfw\|\right\}, \\
		-\|\bfw\| &\leq \left| \|\tau \bfmu +\bfw\|-\tau   \right| &\leq  2\tau  -\|\bfw\| \qquad \mbox{when}\quad \|\tau \bfmu +\bfw\|\leq \tau \leq \|\bfw\|.
	\end{eqnarray*}
	Finally, when $\tau <\|\bfw\|$ then $\|\bfw\|-2\tau\geq -\|\bfw\|$, and when $2\tau -\|\bfw\|<\|\bfw\|$. Thus, $-\|\bfw\| \leq \left| \|\tau \bfmu +\bfw\|-\tau   \right| \leq  \|\bfw\|$ for all choices of $\tau $ and $\bfw$. The result for $d>2$ immediately follows by inserting the above inequality in the kernel.
	
	For $d=2$, i.e., $(d/2-1)=0$, observe that for $x\geq 0$, $
	I_0(x)=\frac{1}{\pi} \int_0^{\pi} \exp\left( x \mathrm{cos}\theta \right) d\theta \implies 
	e^{-x} \leq I_0(x)\leq e^{x}$. 	Therefore,
	\begin{eqnarray*}
		\exp\left(-2\tau-\|w\| \right)	\leq \exp\left(-\|\tau \bfmu +\bfw\|-\tau  \right)
		&\leq & \displaystyle\frac{(\tau \bfmu + \bfw)^{-(d/2-1)} I_{d/2-1}(\tau \bfmu +\bfw)}{(\tau)^{-(d/2-1)} I_{d/2-1}(\tau)} \\
		 &\leq&  \exp \left(\|\tau \bfmu +\bfw\|+\tau \right) \leq \exp\left(2\tau +\|\bfw\| \right). 
	\end{eqnarray*}
	This completes the proof for $d=2$.
\end{proof}

\subsection{Proof of Theorem \ref{thm:1}}\label{subsec:A2}
\begin{proof}
	The proof uses ideas of similar proofs from \cite{WG2008}.
	The proof is done in three parts shown in the following lemma, which is same as Theorem 1 in \cite{WG2008}. 
	\begin{lemma}\label{lm:2}
		Let $\sigma^2\in S$, $\bta\in \Theta$, $P$ be the mixing distribution on $\Theta$, i.e., $f_{P,\sigma^2}(\cdot)=\int \mathrm{FG}_{\sigma} (\cdot\mid\Theta) dP(\Theta) $ where $\mathrm{FG}_{\sigma} (\cdot\mid\Theta)$ is as described in (\ref{FG}). Further, let $\mathcal{M}\left(\Theta \right)$ be the space of probability measures on $\Theta$. 
		Suppose there exists $P_{\varepsilon}$, $\sigma^2_{\varepsilon}$, $S\in \mathbb{R}^{+}$ and $\mathcal{W}\in \mathcal{M}\left(\Theta \right)$, such that $p(\mathcal{W})>0$ and $p(S)>0$, where $p$ denotes the prior distribution, and the followings hold:
		\begin{enumerate}[I.]
			\item $\displaystyle\int_{\RR^d} f_0(\bfx) \log \displaystyle\frac{f_0(\bfx)}{f_{P_\varepsilon,\sigma_\varepsilon^2}(\bfx)}d\bfx <\varepsilon$,
			\item $\displaystyle\int_{\RR^d} f_0(\bfx) \log \displaystyle\frac{f_{P_\varepsilon,\sigma_\varepsilon^2}(\bfx)}{f_{P_\varepsilon,\sigma^2}(\bfx)}d\bfx <\varepsilon$ for every $\sigma^2\in S$, 
			\item $\displaystyle\int_{\RR^d} f_0 \log \displaystyle\frac{f_{P_\varepsilon,\sigma^2}(\bfx)}{f_{P,\sigma^2}(\bfx)}d\bfx <\varepsilon$ for every $P\in\mathcal{W}$ and $\sigma^2\in S$,
		\end{enumerate}
		then $f_0\in KL\left(\Pi\right)$, where $KL(\Pi)$ is as described in Section \ref{sec:3}.
	\end{lemma}
	\noindent The proof is same as the proof of \cite[Theorem 1]{WG2008}, as $\sigma^2$ is independent of $P$.
	
	The proof of Theorem \ref{thm:1} relies on showing that the above conditions hold for our proposed prior. Without loss of generality we assume that the fixed point $\bfx_0$ in assumption C is ${\bf 0}$.  
	
	\noindent{\it Showing condition (I) holds:} We choose $P_{\varepsilon}$ and $\sigma_{\varepsilon}^2$ as follows. For any $\varepsilon>0$ we fix an $m_{\varepsilon}$ such that
	\begin{eqnarray*}
		f_{\varepsilon}(\bfc)=\begin{cases}
			t_m f_0(\bfc)&\|\bfc\|<m_{\varepsilon},\\
			0 & \mbox{otherwise,}
		\end{cases}
	\end{eqnarray*}
	where $t_m^{-1}=\int_{\|\bfc\|<m_{\varepsilon}} f_0(\bfc)dc$,
	and $r=m_{\varepsilon}^{-\eta}$ for some $\eta>0$ with probability 1. The weight $\bfpi_\varepsilon=\bfpi_0$ for any $\bfpi_0\in \Delta_M$, $\Delta_M$ being the $M$-dimensional simplex, with probability 1. Let $F_\varepsilon$ be the probability measure corresponding to $f_\varepsilon$, then $P_{\varepsilon}=F_\varepsilon\times \delta\left(m_{\varepsilon}^{-\eta}\right)$,
	where $\delta$ denotes the degenerate distribution. Also let $\sigma_\varepsilon=r_\varepsilon=m_{\varepsilon}^{-\eta}$, with probability 1. 
	Here $(m_\varepsilon,\eta)$ is such that $E\left(\|\bfx\|^{2(\eta+1)}\right)<\infty$ (see assumption C), and $m_\varepsilon^{2\eta}>4s$, where $s$ is as in assumption B.
	
	Therefore, $f_\varepsilon$ is as follows:
	\begin{eqnarray*}
		f_{\varepsilon}(\bfx)=\int_{\|\bfc\|<m_\varepsilon} \sum_{j=1}^{M} \pi_{0,j} \frac{t_m C_d(\tau_j)\left(2\pi\sigma_\varepsilon^2\right)^{-(d/2)}}{C_d\left(\|\tau_j \bfmu_j + r_\varepsilon (\bfx-\bfc)/\sigma_\varepsilon^2\|\right)} 
		\exp\left\{ -\frac{1}{2} \left(\frac{\|\bfx-\bfc\|^2}{\sigma_\varepsilon^2}+1\right)\right\} f_0(\bfc) d\bfc.
	\end{eqnarray*}
	Next taking a transformation $(\bfc-\bfx)/\sigma_\varepsilon=\bfw$, we get $f_{\varepsilon}(\bfx)$ as follows
	$$ \int_{\RR^d} I\left(\|\bfx+\sigma_\varepsilon \bfw\|<m_\varepsilon\right) \sum_{j=1}^{M} \pi_{0,j} \frac{t_m C_d(\tau_j)\left(2\pi\right)^{-(d/2)}}{C_d\left(\|\tau_j\bfmu_j+\bfw\|\right)}  \exp\left\{ - \frac{1}{2} \left(\|\bfw\|^2+1\right)\right\} f_0(\bfx+\sigma_\varepsilon \bfw) d\bfw,$$
	where $I(\cdot)$ is the indicator function.
	Observe that, as $m_\varepsilon \rightarrow \infty$, $t_m\rightarrow 1$ and the integrand above converges to $\sum_{j} \pi_{0,j} h_1\left(\bfw,({\bf 0},1),(\bfmu_j,\tau_j)\right)f_0(\bfx)$.
	
	Observe that $t_m\leq t_1$ for $m>1$, and by Lemma \ref{lm:3} the above integrand is no bigger than
	
	\begin{eqnarray}
	t_1\left(2\pi\right)^{-(d/2)}  \exp\left\{ - \frac{1}{2} \left(\|\bfw\|-1\right)^2  \right\} f_0(\bfx+\sigma_\varepsilon \bfw)  \leq M_1  t_1 \left(2\pi\right)^{-(d/2)}  \exp\left\{ - \frac{1}{2} \left(\|\bfw\|-1\right)^2  \right\},\label{eq_3} 
	\end{eqnarray}
	for $d>2$ and a suitable constant $M_1>0$ due to assumption A.
	The last term is $\bfw$-integrable. Using polar transformation one can show that the above integral is equal to $M_1 t_1 2^{-(d-3)/2}\pi^{-1/2}/\Gamma(d/2) $. 
	For $d=2$, the above integrand is the same as (\ref{eq_3}) with a multiplier $\sum_j \pi_{0,j} \exp\left(2\tau_j\right)\leq \exp\left(2\tau_{\max} \right)$, and therefore $\bfw$-integrable. Here $\tau_{\max}=\max\{ \tau_1,\ldots, \tau_M \}$. Therefore by the dominated convergence theorem (DCT), $f_{\varepsilon}(\bfx)\rightarrow f_0(\bfx)$, as $m_\varepsilon\rightarrow\infty$.
	
	Next we show that $\left|f_0(\bfx) \log\left(f_0(\bfx)/f_\varepsilon(\bfx) \right)\right|$ is bounded by an integrable function so that the DCT can be applied to the integrand in (I). Observe that
	\begin{align}
	\frac{f_{\varepsilon}(\bfx)}{f_0(\bfx)}&=\int_{\|\bfc\|<m_\varepsilon} \sum_{j=1}^{M} \pi_{0,j} \frac{t_m C_d(\tau_j)\left(2\pi\sigma_\varepsilon^2\right)^{-(d/2)}}{f_0(\bfx) C_d\left(\|\tau_j \bfmu_j + r_\varepsilon(\bfx-\bfc)/\sigma_\varepsilon^2\|\right)} 
	\exp\left\{ -\frac{1}{2} \left(\frac{\|\bfx-\bfc\|^2}{\sigma_\varepsilon^2}+1\right)\right\} f_0(\bfc) d\bfc
	\notag \\
	&\leq \int_{ \mathbb{R}^d} \sum_{j=1}^{M} \pi_{0,j} \frac{\exp\left(2\tau_j \right)t_m \left(2\pi\sigma_\varepsilon^2\right)^{-(d/2)}}{f_0(\bfx)}
	\exp\left\{-\frac{1}{2}  \left( \frac{\|\bfx-\bfc\|}{\sigma_\varepsilon}-1 \right)^2  \right\} f_0(\bfc) d\bfc \notag\\
	&\leq \int_{\mathbb{R}^d} \frac{\exp\left( \tau_{\max}\right)t_m \left(2\pi\right)^{-(d/2)}}{f_0(\bfx)}
	\exp\left\{-\frac{1}{2}  \left( \left\|\bfw\right\|-1 \right)^2  \right\} f_0(\bfx+\sigma_\varepsilon \bfw) d\bfw \qquad \notag\\
	&\leq \int_{ \mathbb{R}^d} \frac{\exp\left(2\tau_{\max}\right)t_m \left(2\pi\right)^{-(d/2)}}{f_0(\bfx)}
	\exp\left\{-\frac{1}{2}  \left( \left\|\bfw\right\|-1 \right)^2 \right\} \notag\\
	& \hspace{2.25 in} \left\{f_0(\bfx)+\sigma_\varepsilon^{r_0} \| \bfw\|^{r_0} L(\bfx)\exp\left(s \sigma_\varepsilon^2 \|\bfw\|^2  \right) \right\} d\bfw, 
	\end{align}
	where the inequality in the second line holds by Lemma \ref{lm:3}, and that in the last line by assumption B. Without loss of generality assume that $m_{\varepsilon}>M^{*}$ for some $M^{*}$ such that $ \exp\{2\tau_{\max}\}t_m<t_0$ for some positive constant $t_0$. Therefore the above integral reduces to
	\begin{eqnarray*}
		&& \frac{f_{\varepsilon}(\bfx)}{f_0(\bfx)}\leq t_0 \left(2\pi\right)^{-(d/2)} \left[ \int_{ \RR^d} 
		\exp\left\{-\frac{1}{2}  \left( \left\|\bfw\right\|-1 \right)^2 \right\}d\bfw \right. \\
		&&\hspace{1 in}\left. + \frac{L(\bfx)}{f_0(\bfx)}\int_{\RR^d} 
		\left(\sigma_\varepsilon \| \bfw\|  \right)^{r_0} \exp\left\{-\frac{1}{2}  \left( \left\|\bfw\right\|-1 \right)^2+s \sigma_\varepsilon^2 \|\bfw\|^2  \right\}d\bfw \right].
	\end{eqnarray*}
	Observe that both the integrals are finite as long as $s\sigma_\varepsilon^2<1/3$ (i.e., $s<m^{2\eta}/3)$. To see this, observe that $\sigma_\varepsilon<1$, and $s\sigma_\varepsilon^2\eqqcolon \xi<1/3$. Hence the second integral is no bigger than
	\begin{eqnarray*}
		&& \int_{\RR^d} \| \bfw\|^r \exp\left\{\frac{2\xi-1}{2}  \left( \left\|\bfw\right\|-\frac{1}{1-2\xi}\right)^2-\frac{1}{2}\left( 1-\frac{1}{1-2\xi}\right) \right\}d\bfw\\
		&& \hspace{1 in}< \exp\left(\frac{1}{4}\right)\int_{\RR^d} \| \bfw\|^r \exp\left\{\frac{2\xi-1}{2}  \left( \left\|\bfw\right\|-\frac{1}{1-2\xi}\right)^2 \right\}d\bfw.
	\end{eqnarray*}
	Now it is easy to see that the above integral is finite for $\xi<1/3$. In particular, one can use polar transformation to evaluate the same. Similarly one can also show that the first integral is also finite. Therefore
	\begin{eqnarray}
	\int_{\RR^d} f_0(\bfx) \log \frac{f_\varepsilon(\bfx)}{f_0(\bfx)}d\bfx &\leq& \int_{\RR^d} f_0(\bfx) \log \left\{ M_0+M_1\frac{L(\bfx)}{f_0(\bfx)}\right\}d\bfx \notag \\  
	&\leq& \log M_0 +\frac{M_1}{M_0} E_0\left\{\frac{L(\bfx)}{f_0(\bfx)}\right\}  <\infty, \label{eq_6}
	\end{eqnarray}
	as $\log(1+u)<u$ and $E_0\left\{L(\bfx)/f_0(\bfx)\right\}<\infty$, i.e., $L$ is integrable by assumption B.
	
	Next we find a lower-bound $g(\bfx)$ of $f_\varepsilon(\bfx)$. Towards that from 
	\cite{MBFbound}, $y^{-\nu} I_\nu(y)$ is a strictly increasing function over $(0,\infty)$, and therefore $y^{\nu} I^{-1}_\nu(y)$ is strictly decreasing over $(0,\infty)$. Therefore $C_d(\tau_0)/C_d(\tau)>1$ for any $(\tau,\tau_0)$ such that $\tau>\tau_0$. For $\tau<\tau_0$, from \citet[Equation (2.2)]{MBFbound} we have	$C_d(\tau_0)/C_d(\tau)> \exp\left(\tau-\tau_0\right) >\exp\left(-\tau_0\right)$.
	Therefore,
	$$f_\varepsilon(\bfx) \geq \exp\left(-\tau_{\min}-0.5\right)(2\pi \sigma_\varepsilon^2)^{-d/2} \int_{\|\bfc\|<m_\varepsilon} t_m \exp\left\{ -\frac{1}{2\sigma_\varepsilon^2}\left(\|\bfx\|+\|\bfc\| \right)^2\right\} f_0(\bfc) d\bfc. $$
	Now let $\|\bfx\|\geq m_\varepsilon/2$. Then the above expression is no less than
	\begin{align*}
	f_\varepsilon(\bfx) &\geq \exp\left(-\tau_{\min}-0.5\right) (2\pi\sigma_\varepsilon^2)^{-d/2}\exp\left( -\frac{9}{2\sigma_\varepsilon^2} \|\bfx\|^2\right) t_m \int_{\|\bfc\|<m_\varepsilon} f_0(\bfc) d\bfc \\
	&= \exp\left(-\tau_{\min} -0.5\right)(2\pi\sigma_\varepsilon^2)^{-d/2}\exp\left( -\frac{9}{2\sigma_\varepsilon^2} \|\bfx\|^2\right) \eqqcolon g(m_\varepsilon).
	\end{align*}
	Note that $g(m_\varepsilon)$ is a decreasing function of $m_\varepsilon$ for sufficiently large $m_\varepsilon$. When $m_\varepsilon$ is large, then $g(m_\varepsilon)\leq g(2\|\bfx\|)$, i.e., the above expression is no less than
	\begin{eqnarray*}
		(2\pi)^{-d/2}(2\|\bfx\|)^{d\eta}\exp\left\{ -9\times 2^{2\eta-1} \|\bfx\|^{2(\eta+1)}-\left(\tau_{\min} +\frac{1}{2}\right)\right\} =K_1 \|\bfx\|^{d\eta} \exp\left\{-K_2 \|\bfx\|^{2(\eta+1)} \right\},
	\end{eqnarray*}
	for some suitable constants $K_1$ and $K_2$.
	
	Let $\|\bfx\|<m_\varepsilon/2$, $0<\delta<m_\varepsilon/3$ fixed, and $\phi_m(\bfx)=\inf_{\|{\bf t}- \bfx \|<\delta\sigma_\varepsilon} f_0({\bf t})$, then $\{\bfc:\|\bfc\|<m_\varepsilon\}\cap\{\|\bfc-\bfx\|\leq \delta \sigma^2_\varepsilon \}=\{\bfc: \|\bfc-\bfx\|\leq \delta \sigma^2_\varepsilon \}$ for some small enough $\delta>0$ and
	
	\begin{align*}
	f_\varepsilon(\bfx) &\geq \int_{\|\bfc-\bfx\|\leq \delta \sigma^2_\varepsilon } t_m \exp\left(-\tau_{\min} -0.5\right) (2\pi\sigma_\varepsilon^2)^{-d/2}
	\exp\left\{ -\frac{1}{2} \left(\frac{\|\bfx-\bfc\|^2}{\sigma_\varepsilon^2}+1\right)\right\} f_0(\bfc) d\bfc\\
	&\geq t_m \exp\left\{-(\tau_{\min}+1)\right\} (2\pi\sigma_\varepsilon^2)^{-d/2}  \int_{\|\bfc-\bfx\|\leq \delta \sigma^2_\varepsilon } 
	\exp\left( -\frac{\|\bfx-\bfc\|^2}{2\sigma_\varepsilon^2}\right) f_0\left(\frac{\bfc-\bfx}{\sigma_\varepsilon}\sigma_\varepsilon+\bfx \right) d\bfc\\
	&\geq t_m \exp\left\{-(\tau_{\min}+1)\right\} (2\pi)^{-d/2} \int_{\|\bfw\|\leq \delta } \exp\left( -\frac{\|\bfw\|^2}{2}\right) f_0(\bfw\sigma_\varepsilon+\bfx) d\bfw\\
	&\geq t_m \exp\left\{-(\tau_{\min}+1)\right\} (2\pi)^{-d/2}  \int_{\|\bfw\|\leq \delta } \exp\left(-\frac{\|\bfw\|^2}{2}\right)\\
	&\hspace{2.25 in} \left\{f_0(\bfx)-L(\bfx) \exp\left(s\sigma_\varepsilon^2 \|\bfw\|^2 \right) \|\bfw\sigma_\varepsilon\|^{r_0}\right\} d\bfw\\
	&\geq t_m \exp\left\{-(\tau_{\min}+1)\right\} \left[ f_0(\bfx)   P\left\{\|\bfw\| \leq \delta \mid \bfw \sim N({\bf 0},I)\right\} \right. \\ 
	& \hspace{1.25 in}\left. -L(\bfx)   \sigma_\varepsilon^{r_0}(1-2s\sigma_\varepsilon^2)^{-(r_0 +d)/2} \int_{\|\bfu\|<\delta\sqrt{1-2s\sigma_\varepsilon^2}} \|\bfu\|^{r_0}\phi(\bfu)d\bfu\right].
	\end{align*}
	We bound the last term inside the bracket by 
	$$ L(\bfx)   \frac{\left(\delta\sigma_\varepsilon\right)^{r_0}}{(1-2s\sigma_\varepsilon^2)^{(r_0+d)/2}} \left(1-2s\sigma_\varepsilon^2\right)^{r_0/2} P\left\{\|\bfw\|< \delta\sqrt{1-2s\sigma_\varepsilon^2} \mid \bfw\sim N({\bf 0},I) \right\}. $$
	Let $\sigma_\varepsilon$ be sufficiently small (i.e., $m_\varepsilon$ be sufficiently large) such that $0< s\sigma_\varepsilon^2 < 1/4$, which implies that $1/2<1-2s\sigma_\varepsilon^2<1$. Then the above term is no bigger than
	$$L(\bfx)  2^{d/2} \left(\delta\sigma_\varepsilon\right)^{r_0}  P\left(\|\bfw\|< \delta\sqrt{1-2s\sigma_\varepsilon^2} \mid \bfw\sim N({\bf 0},I) \right). $$
	Combining the above, noting that $t_m\geq1$ and by assumption B. we get:
	\begin{eqnarray*}
		f_\varepsilon(\bfx) \geq \exp\left\{-(\tau_{\min} +1)\right\}  f_0(\bfx)   \left[  P\left\{\|\bfw\| \leq \delta\mid \bfw \sim N({\bf 0},I)\right\} - M 2^{d/2} \left(\delta\sigma_\varepsilon\right)^{r_0}  \right.\\
		\qquad \left. P\left\{\|\bfw\|< \delta\sqrt{1-2s\sigma_\varepsilon^2} \mid\bfw\sim N({\bf 0},I) \right\} \right] \geq K^{*} f_0(\bfx), 
	\end{eqnarray*}
	for some suitable constant $0<K^{*}<1$. The last inequality is true of we take $m_\varepsilon$ large enough and $\delta$ small enough. Therefore
	$$f_\varepsilon(\bfx) \geq\begin{cases}
	K^{*}f_0(\bfx) & \|\bfx\|<R,\\
	K^{*}f_0(\bfx)\wedge \left\{ K_1 \|\bfx\|^{d\eta} \exp\left( -K_2\|\bfx\|^{2(\eta+1)}\right)\right\} & \|\bfx\|\geq R,
	\end{cases}$$
	for any $R<m_\varepsilon/2$. Consequently,
	$$\log \frac{f_0(\bfx)}{f_\varepsilon(\bfx)} \leq \xi(\bfx)\coloneqq \begin{cases}
	-\log K^{*} & \|\bfx\|<R,\\
	-\log K^{*}\vee \left\{\log\frac{\displaystyle f_0(\bfx)}{ \displaystyle K_1 \|\bfx\|^{d\eta} \displaystyle\exp\left( -K_2\|\bfx\|^{2(\eta+1)}\right)}\right\}& \|\bfx\|\geq R.
	\end{cases}$$
	Recalling that $1/K^{*}\geq 1$, we have
	\begin{eqnarray}
	\int_{\RR^d} f_0(\bfx) \xi(\bfx) d\bfx \leq
	-\log K^{*}+\int_{A} f_0(x) \log\frac{f_0(\bfx)}{ K_1 \|\bfx\|^{d\eta} \exp\left( -K_2\|\bfx\|^{2(\eta+1)}\right)} d\bfx, \label{eq_4}
	\end{eqnarray}
	where $A\coloneqq\left\{\bfx: \|\bfx\|\geq R,~~\mbox{and}~~ f_0(\bfx) \geq K_1 \|\bfx\|^{d\eta} \exp\left( -K_2\|\bfx\|^{2(\eta+1)} \right) \right\}$. 
	To see that the second term above is finite, observe that as
	$f_0(\bfx) \geq K_1 \|\bfx\|^{d\eta} \exp\left( -K_2\|\bfx\|^{2(\eta+1)} \right)\geq 0$ for all $\bfx$ in $A$, therefore
	\begin{eqnarray}
	f_0(\bfx) \log f_0(\bfx) \geq f_0(\bfx) \log \left\{K_1 \|\bfx\|^{d\eta} \exp\left( -K_2\|\bfx\|^{2(\eta+1)}\right) \right\}\quad  \mbox{for all $\bfx$  in $A$,} \notag \\
	\implies  \int_{A} f_0(\bfx) \log f_0(\bfx) d\bfx \geq \int_{A} f_0(\bfx) \log \left\{K_1 \|\bfx\|^{d\eta} \exp\left( -K_2\|\bfx\|^{2(\eta+1)}\right) \right\} d\bfx. \label{eq_5}
	\end{eqnarray}
	If we show that the RHS of (\ref{eq_5}) has a lower-bound and the LHS of (\ref{eq_5}) has an upper-bound then the second term of (\ref{eq_4}) is also finite. Towards that,
	\begin{eqnarray*}
		&&\int_{A} f_0(\bfx) \log \left\{K_1 \|\bfx\|^{d\eta} \exp\left( -K_2\|\bfx\|^{2(\eta+1)}\right) \right\} d\bfx \\
		&& = d\eta \int_{A}  \log (K_3 \|\bfx\|)f_0(\bfx) d\bfx- K_2 \int_{\bfx\in A} \|\bfx\|^{2(\eta+1)} f_0(\bfx) d\bfx\\
		&&\geq d\eta \log (K_3 R) \int_{A} f_0(\bfx) d\bfx - K_2 E\left(\|\bfx\|^{2(\eta+1)} \right) \geq -C
	\end{eqnarray*}
	for some constant $C>0$, $K_3=K_1/d\eta$ as $E\left(\|\bfx\|^{2(\eta+1)} \right)<\infty$, and $\|\bfx\|\geq R$ in $A$. Again $\int f_0(\bfx) \log(f_0)d\bfx< \log M$ as $f_0(\bfx)<M$ for all $\bfx\in \mathbb{R}^d$. Thus (\ref{eq_4}) is finite. 
	
	As $|a|=\max\{a,-a\}$, from (\ref{eq_6}) and the above arguments $f_0\log(f_0/f_\varepsilon) \leq |f_0\log(f_0/f_\varepsilon)|$ and $\int |f_0\log(f_0/f_\varepsilon)|$ is finite, and therefore the DCT holds. Thus (I) is satisfied.
	
	\vskip10pt
	\noindent{\it Showing condition (II) holds:}  The proof follows in 4 steps:	
	\begin{enumerate}[i.]
		\item We will first show that $h_\sigma\left\{\bfx,(\bfc,r)\right\}\rightarrow h_{\sigma_\varepsilon} \left\{\bfx, (\bfc,r) \right\}$ for any given $\bfx$ and $(\bfc^{\prime},r)^{\prime}$, as $\sigma \rightarrow \sigma_\varepsilon$.
		\item We use the DCT to show that $f_{P_\varepsilon,\sigma}\rightarrow f_{P_\varepsilon,\sigma_\varepsilon}$ for $\sigma$ in the neighborhood of $\sigma_\varepsilon$, $\sigma\in N(\sigma_\varepsilon)$, by choosing an appropriate dominating function.
		\item Then we apply (i) to show that $ \displaystyle\int_{\RR^d} f_0(\bfx) \log \displaystyle\frac{f_{P_{\varepsilon},\sigma_\varepsilon}(\bfx)}{f_{P_{\varepsilon},\sigma}(\bfx)}d\bfx  \rightarrow 0$  using DCT as $\sigma\rightarrow \sigma_\varepsilon$, by finding an appropriate dominating function on $\bta\in D$ where $\{(\bfc,r,\bfpi_0): \|\bfc\|<m_\varepsilon, r=m_\varepsilon^{-\eta}\}\subseteq D$.
		\item The proof is completed by taking $S=\{\sigma: |\sigma-\sigma_\varepsilon|<\delta\}\cap N(\sigma_\varepsilon)$, showing that $\pi(S)>0$.
	\end{enumerate}
	
	The proof of (i) follows from the fact that $h_\sigma\left\{\bfx,(\bfc,r)\right\}$ is a (pointwise) continuous function of $\sigma$ on $(0,\infty)$ given $\bfx$ and $(\bfc,r)$. To see this observe that $\sigma^{-d}$, $\exp\left\{ -\left(\|\bfx-\bfc\|^2+r^2\right)/\left(2\sigma^2\right)\right\}$ and $\|\tau\bfmu_0+r(\bfx-\bfc)/\sigma^2\|$ are continuous in $(0,\infty)$. Further, for any fixed $\tau\in\{\tau_1,\ldots,\tau_M \}$, $C_d^{-1}(\tau)\leq C_d^{-1}(\tau_{\max})\approx (2\pi)^{d/2} \sum_{m=0}^{\infty} (\tau_{\max}/2)^{2m}/\left\{m! \Gamma(d/2+m) \right\}$ is finite. These together with the facts that product and convolution of continuous functions are continuous, shows that $h_\sigma\left\{\bfx,(\bfc,r)\right\}\rightarrow h_{\sigma_\varepsilon}\left\{\bfx,(\bfc,r)\right\}$ as $\sigma\rightarrow\sigma_\varepsilon$ pointwise.  
	
	To show (ii) we first fix a neighborhood of $\sigma_\varepsilon$. Recall that $\sigma_\varepsilon=m_\varepsilon^{-\eta}$. We fix the neighborhood as $N(\sigma_{\varepsilon})=( \underline{\sigma},\bar{\sigma})$, such that $\underline{\sigma}<\sigma_\varepsilon<\bar{\sigma}$. 
	Next we define $g_\sigma\left\{\bfx,(\bfc,r)\right\}=(2\pi \sigma^2)^{-d/2} \exp\left\{- \left(\|\bfx-\bfc\| - r\right)^2/(2\sigma^2) \right\}$.
	Note that $g_\sigma\left\{\bfx,(\bfc,r)\right\}$ is an increasing function of $\sigma$ if $\sigma<\left|\|\bfx-\bfc\|-r \right|/\sqrt{d}$. To see this consider the following:
	$$\frac{\partial g_\sigma\left\{\bfx,(\bfc,r)\right\}}{\partial \sigma}=(2\pi)^{-d/2} \sigma^{-(d+1)} \left\{\frac{(\|\bfx-\bfc\|-r)^2}{\sigma^2}-d\right\}   \exp\left\{- \frac{\left(\|\bfx-\bfc\| - r\right)^2}{2\sigma^2} \right\}. $$
	Let $A=\{(\bfc,r): |\|\bfx-\bfc\|-r|>\sqrt{d} \bar{\sigma} \}$, then $h_{\sigma}\left\{\bfx,(\bfc,r)\right\}\leq e^{2\tau_{\max}}g_{\bar{\sigma}}\left\{\bfx,(\bfc,r)\right\}$ for $(\bfc,r) \in A$ by Lemma \ref{lm:3}.
	For $A^c=\{(\bfc,r): |\|\bfx-\bfc\|-r|<\sqrt{d} \bar{\sigma} \}$ we bound the kernel $h_{\sigma}\left\{\bfx,(\bfc,r) \right\}$ by
	$ g_{\underline{\sigma}}\left\{\bfx,(\bfc,r)\right\}=(2\pi \underline{\sigma}^2)^{-d/2} \exp\left\{- \left(\|\bfx-\bfc\| - r\right)^2/(2\bar{\sigma}^2) \right\}$. Now it is easy to see that $g_{\underline{\sigma}}\left\{\bfx,(\bfc,r)\right\}$ and $g_{\bar{\sigma}}\left\{\bfx,(\bfc,r)\right\}$ are $P_{\bfc,r}$-integrable and $g(\bfx,\bta)=\max\left[ g^{\prime}\left\{\bfx,(\bfc,r)\right\},g_{\bar{\sigma}}\left\{\bfx,(\bfc,r)\right\}\right]<C_0$, for some appropriate constant $C_0$, depending on $(\underline{\sigma},\bar{\sigma})$. Thus (ii) holds.

	Define $D=\{(\bfc,r,\bfpi): \|c\|<R, \underline{r}<r<\bar{r}; \bfpi\in \Delta_M \}$. To prove (iii) we will show that for any $\sigma\in N(\sigma_\varepsilon)$:
	\begin{eqnarray*}
	\int_{\RR^d} f_0(\bfx) \left|\log \left\{ \sup_{\bta\in D}  \sum_j \pi_j^{(\bfc,r)} h_\sigma(\bfx,\bta) \right\} \right| d\bfx <\infty  \quad\\
	 \mathrm{and}\quad
	\int_{\RR^d} f_0(\bfx) \left|\log \left\{ \inf_{\bta\in D}  \sum_j \pi_j^{(\bfc,r)} h_\sigma(\bfx,\bta) \right\} \right| d\bfx <\infty  .
	\end{eqnarray*}
	For any $\sigma\in N(\sigma_\varepsilon)$, and by Lemma \ref{lm:3} 
	\begin{eqnarray*}
		\left| \log \sup_{\bta\in D} \sum_j \pi_j^{(\bfc,r)} h_\sigma(\bfx,\bta)\right|  \leq \int_{\RR^d} f_0 \left| \log \left\{ \sup_{\bta\in D} \sum_j \pi_j^{(\bfc,r)} h_\sigma(\bfx,\bta) \right\}\right| d\bfx. 
	\end{eqnarray*}
	Now $ \sup_{\bta\in D} \sum_j \pi_j^{(\bfc,r)} h_\sigma(\bfx,\bta)\leq \sup_{\bfpi \in \Delta_M} \sum_j \pi_j^{(\bfc,r)} \sup_{(\bfc,r)\in D} h_\sigma(\bfx,\bta)  $. Note that $\sup_{(\bfc,r)\in D} h_\sigma(\bfx,\bta)  $ is no bigger than
	\begin{eqnarray*}
		\exp\left\{	 \left|2\tau_{\max}-\frac{d}{2}\log \left(2\pi \sigma^2\right) -\frac{1}{2\sigma^2} \inf_{\bta\in D} \left(\|\bfx-\bfc\|-r \right)^2 \right| \right\}  \leq \exp\left\{ 2\tau_{\max}+ \frac{d}{2}\log \left(2\pi \sigma^2\right) \right\} .
	\end{eqnarray*}
	and therefore
	\[\log \sup_{\bfpi \in \Delta_M} \sum_j \pi_j^{(\bfc,r)} \sup_{(\bfc,r)\in D} h_\sigma(\bfx,\bta) \leq  2\tau_{\max}+ \frac{d}{2}\log \left(2\pi \sigma^2\right) , \]
	which is $f_0$-integrable. Similarly $	\left| \log \inf_{\bta\in D} \sum_j \pi_j h_\sigma(\bfx,\bta)\right|$ is no bigger than
	\begin{eqnarray}
	&& \left|\log \inf_{\bfpi\in \Delta_M } \sum_j \pi_j^{(\bfc,r)}  \exp \left\{\inf_{(\bfc, r)\in D}  \log h_\sigma(\bfx,\bta) \right\} \right| \notag \\
	&&\leq \left|\log \inf_{\bfpi\in \Delta_M } \sum_j \pi_j^{(\bfc,r)}  \exp \left\{ -2\tau_{\min} - \frac{d}{2} \log \left(2\pi \sigma^2\right) -\frac{1}{2\sigma^2} \sup_{\bta\in D} \left(\|\bfx-\bfc\|+r \right)^2 \right\} \right| \notag \\
	&&\leq  \left|\log \inf_{\bfpi\in \Delta_M } \sum_j \pi_j^{(\bfc,r)}  \exp  \left[ -2\tau_{\min} -\frac{d}{2}\log \left(2\pi \sigma^2\right) \right. \right. \notag \\
	&&\hspace{1.25 in} \left. \left. -\frac{1}{2\sigma^2}  \left\{\|\bfx\|^2+R^2+\bar{r}^2+2\bar{r}( \|\bfx\|+R) \right\} \right] \right| \notag \\
	&&= \left|-2\tau_{\min} -\frac{d}{2}\log \left(2\pi \sigma^2\right) -\frac{1}{2\sigma^2}  \left\{\|\bfx\|^2+R^2+\bar{r}^2+2\bar{r}( \|\bfx\|+R) \right\} \right|.  \label{eq_9}
	\end{eqnarray}
	The last expression is also $f_0$-integrable as $E_0\left(\|\bfx\|^2\right)<\infty$. Thus we dominate 
	\begin{eqnarray*}
		\left| \log \frac{f_{P_\varepsilon},\sigma_\varepsilon}{f_{P_\varepsilon},\sigma} \right| \leq \max\left\{\left|\log \frac{\sup_{\bta\in D} \sum_j \pi_j h_{\sigma_\varepsilon}(\bfx,\bta)}{\inf_{\bta\in D} \sum_j \pi_j h_{\sigma}(\bfx,\bta)} \right|, \left|\log  \frac{\sup_{\bta\in D}  \sum_j \pi_j h_{\sigma}(\bfx,\bta)}{\inf_{\bta\in D} \sum_j \pi_j h_{\sigma_\varepsilon}(\bfx,\bta)} \right| \right\}\\
		\leq d\log \left(\frac{\sigma}{\sigma_\varepsilon}\vee \frac{\sigma_\varepsilon}{\sigma} \right)+\frac{1}{2 (\sigma^2 \wedge \sigma_\varepsilon^2)} \left\{\|\bfx\|^2+R^2+\bar{r}^2+2R\bar{r}( \|\bfx\|+R) \right\},
	\end{eqnarray*}
	which is $f_0$-integrable for $\sigma\in N(\sigma_\varepsilon)$. Therefore by the DCT 
	$$\int_{\RR^d} f_0(\bfx)\log \left\{ f_{P_\varepsilon,\sigma_\varepsilon}(\bfx)/f_{P_\varepsilon,\sigma}(\bfx)\right\}d\bfx \rightarrow 0 . $$
	
	Therefore,  for any given $\varepsilon>0$, there exists $\delta>0$ such that 
	$$\int_{\RR^d} f_0(\bfx)\log \left\{ f_{P_\varepsilon,\sigma_\varepsilon}(\bfx)/f_{P_\varepsilon,\sigma}(\bfx)\right\}d\bfx<\varepsilon$$ whenever $|\sigma-\sigma_\varepsilon|<\delta$. Thus we choose $S=\left\{\sigma:\underline{\sigma},\bar{\sigma} \right\}\cap \left\{\sigma:\sigma_\varepsilon-\delta,\sigma_\varepsilon+\delta \right\}= \left\{\sigma:\underline{\sigma}^{*},\bar{\sigma}^{*} \right\}$, say. The proof of (II) is complete noticing that $\pi(S)>0$, $S$ is nonempty (it contains a neighborhood of $\sigma_\varepsilon$) as $\sigma^2$ follows an inverse-gamma prior.
	
	\vskip10pt
	\noindent{\it Showing condition (III) holds:} To check the last condition we apply the following Lemma 3 of \cite{WG2008}, which states that (III) holds if the following three conditions hold:
	\begin{enumerate}[i.]
		\item For any $\sigma\in S$, $\displaystyle{\int_{\RR^d} \left| \log \frac{f_{P_\varepsilon,\sigma}(\bfx)}{\inf_{\bta\in D} \sum_j \pi_j^{(\bfc,r)} h_{\sigma}(\bfx,\bta)} \right| f_0(\bfx) d\bfx<\infty}$.
		\item  Define $c^*\coloneqq\inf_{\bfx\in C} \inf_{\bta\in D} \sum_j \pi_j^{(\bfc,r)} h_{\sigma} (\bfx, \bta)$ then $c^* >0$, for any compact $C\in \mathcal{X}$.
		\item For any $\phi\in S$ and compact $C\in \mathcal{X}$, there exists $E$ containing $D$ in its interior such that the family of maps $\{ \bta \mapsto \sum_j \pi_j^{(\bfc,r)} h_{\sigma}(\bfx,\bta), \bfx\in C \}$ is uniformly equicontinuous on $E\subset \Theta$, and $\sup \left\{ \sum_j \pi_j^{(\bfc,r)}  h_{\sigma} (\bfx,\bta): \bfx\in C, \theta \in E^c \right\} < k \varepsilon /4$, for some $k>0$. 
	\end{enumerate}
	
	To check i., we show that $$\int_{\RR^d} f_0(\bfx) \left| \log f_{P_\varepsilon,\sigma}(\bfx) \right| d \bfx <\infty$$ and $$ \int_{\RR^d} f_0(\bfx) \left| \log \inf_{\theta\in D} \sum_j \pi_j^{(\bfc,r)}  h_{\sigma}(\bfx,\bta)\right| d\bfx <\infty.$$
	 The second part is already shown in the proof of condition (II). To see the first inequality, we proceed in a similar way as in the proof of (I). Observe that for fixed $\sigma\in S$
	\begin{align*}
	f_{P_\varepsilon,\sigma}(\bfx) &= t_m\int_{\|\bfc\|<m_\varepsilon} \sum_j \pi_{0,j}  \frac{C_\sigma C_d(\tau_m)}{C_d \left(\|\bfmu_m\tau_m+ m_\varepsilon^{-\eta}\sigma^{-2}(\bfx-\bfc) \| \right)} \\
	&\hspace{2.5 in} \exp\left\{ -\frac{1}{2\sigma^2} \left(\|\bfx-\bfc\|^2+ m_\varepsilon^{-2\eta} \right)\right\} f_0(\bfc) d\bfc \\
	&\leq  t_m \int_{\|\bfc\|<m_\varepsilon} \sum_j \pi_{0,j}  C_{\sigma} e^{2\tau_m}  \exp\left\{ -\frac{1}{2\sigma^2} \left(\|\bfx-\bfc\|- m_\varepsilon^{-\eta} \right)^2\right\} f_0(\bfc) d\bfc \quad \mbox{[by Lemma \ref{lm:3}]}\\
	&\leq  t_m e^{2\tau_{\min}} C_\sigma \int_{\|\bfc\|<m_\varepsilon} \sum_j \pi_{0,j}   f_0(\bfc) d\bfc =e^{2\tau_{\max}}C_\sigma,
	\end{align*}
	where $C_\sigma=(2\pi\sigma^2)^{-d/2}$. Again observe that
	\begin{eqnarray*}
		f_{P_\varepsilon,\sigma}(\bfx) 
		\geq  t_m \int_{\|\bfc\|<m_\varepsilon} e^{-2\tau_{\min}} C_{\sigma} \sum_j \pi_{0,j}   \exp\left\{ -\frac{1}{2\sigma^2} \left(\|\bfx-\bfc\|+ m_\varepsilon^{-\eta} \right)^2\right\} f_0(\bfc) d\bfc 
	\end{eqnarray*}
	by Lemma \ref{lm:3}. Next note that for $\theta\in D$, $\underline{r}\leq m^{-\eta}_\varepsilon \leq \bar{r}$. Therefore,
	\begin{eqnarray*}
		f_{P_\varepsilon,\sigma}(\bfx)  \geq e^{-2\tau_{\min}} C_{\sigma} t_m \exp \left\{-\frac{1}{\sigma^2}\left(\|\bfx\|^2+ \underline{r}^2\right)\right\}   \int_{\|\bfc\|<m_\varepsilon}\exp\left\{ -\frac{1}{\sigma^2} \|\bfc\|^2\right\} f_0(\bfc) d\bfc.
	\end{eqnarray*}
	By assumption D, there exists some $\delta>0$ such that $\inf_{\|\bfx\|<\delta}f_0(\bfx)=\phi_\delta>0$. Further recall that $t_m\geq 1$, therefore
	\begin{align*}
	f_{P_\varepsilon,\sigma} (\bfx) &\geq e^{-2\tau_{\min}} C_{\sigma} \phi_\delta\exp \left\{-\frac{1}{\sigma^2}\left(\|\bfx\|^2+ \underline{r}^2\right)\right\}   \int_{\|\bfc\|<\delta}\exp\left\{ -\frac{1}{\sigma^2} \|\bfc\|^2\right\}  d\bfc\\
	&\geq e^{-2\tau_{\min}}C_{\sigma} \phi_\delta \exp \left\{-\frac{1}{\sigma^2}\left(\|\bfx\|^2+ \underline{r}^2 \right) \right\}    P\left\{\|\bfc\|< \delta ~\left|~\bfc \sim N({\bf 0},I) \right. \right\}.
	\end{align*}
	For the chosen fixed $\delta>0$ in assumption D, the last term is positive. Therefore (i) follows noting that $E_0\left(\|\bfx\|^2\right)<\infty$.
	
	From (\ref{eq_9}) we get:
	\begin{eqnarray*}
		g(\bfx)= \inf_{\theta\in D} \sum_j \pi_j^{(\bfc,r)} h_\sigma(\bfx,\theta) \geq e^{-2\tau_{\min}}\left(2\pi \sigma^2\right)^{d/2}\exp\left\{ -\frac{1}{2\sigma^2}  \left( \|\bfx\|^2+R+\bar{r}^2+2\bar{r} R+ 2\bar{r} \|\bfx\|  \right) \right\}.
	\end{eqnarray*}
	Note that $g(\bfx)$ is positive, bounded and continuous on $\bfx\in C$, where $C$ is any compact set. Hence (ii) holds.
	
	It remains to show that (iii) is satisfied. Let $C\in\mathcal{X}$ be a given compact set. We need to show that $h_\sigma(\bfx,\bta)$ is uniformly equicontinuous as a family of functions of $\bta=(\bfc,r)$ on a set $E\supset D$. We choose $E=\{(\bfc,r,\bfpi) : \|\bfc\|<a;~ r<\bar{r}^*;~ \bfpi\in \Delta_M \}$ where $0<\bar{r}<\bar{r}^*<\infty$, $a>R$. Clearly, $E$ contains $D$ and is compact. By the definition equicontinuouity, for all $\bta_1=(\bfc_1,r_1,\bfpi_1),\bta_2=(\bfc_2,r_2,\bfpi_2)\in E$ with $\|(\bfc_1,r_1,\bfpi_1)-(\bfc_2,r_2,\bfpi_2)\|<\delta_\epsilon$, we have $\left|\sum_j \pi_{1,j}^{(\bfc_1,r_1)}  h_\sigma(\bfx,\bta_1)- \sum_j \pi_{2,j}^{(\bfc_2,r_2)} h_\sigma(\bfx,\bta_2)\right|<\epsilon$.
	
	For any $\bta_1,\bta_2$,~
	\begin{eqnarray*}
		&& \left|\sum_j \pi_{1,j}^{(\bfc_1,r_1)}  h_\sigma(\bfx,\bta_1)- \sum_j \pi_{2,j}^{(\bfc_2,r_2)} h_\sigma(\bfx,\bta_2) \right|\\
		&& \leq \sum_j \pi_{1,j}^{(\bfc_1,r_1)} \left| h_\sigma(\bfx,\bta_1)- h_\sigma(\bfx,\bta_2) \right|  + \left|\sum_j  \left(\pi_{1,j}^{(\bfc_1,r_1)} -\pi_{2,j}^{(\bfc_2,r_2)}\right)  h_\sigma(\bfx,\bta_2)\right|.
	\end{eqnarray*}
	Thus it is enough to show that $\left| h_\sigma(\bfx,\bta_1)- h_\sigma(\bfx,\bta_2) \right| \leq \epsilon_1/M$ for some appropriate $\epsilon_1< \epsilon$ and  $h_\sigma(\bfx,\bta_2)$ is bounded for $\bfx\in C$ and $\bta \in E$. For the first part, we will show that 
	the ratio  $h_\sigma(\bfx,\bta_1)\left/h_\sigma(\bfx,\bta_2)\right.$ goes to $1$ uniformly, as
	\begin{eqnarray*}
		\frac{h_\sigma(\bfx,\bta_1)}{h_\sigma(\bfx,\bta_2)} \leq\frac{C_d\left(\|\tau_m \bfmu_m +r_2(\bfx-\bfc_2)/\sigma^2\|\right)  }{ C_d\left(\|\tau_m \bfmu_m +r_1(\bfx-\bfc_1)/\sigma^2\|\right)}
		\exp \left[-\frac{1}{\sigma^2}\left\{\|\bfx-\bfc_1\|^2-\|\bfx-\bfc_2\|^2+\left(r_1^2-r_2^{2}\right)\right\} \right].
	\end{eqnarray*}
	Now
	\begin{eqnarray*}
		\left|		\|\bfx-\bfc_1\|^2-\|\bfx-\bfc_2\|^2\right| = 2\left| \|\bfx-\bfc^{*}\| (\bfc_1-\bfc_2)^{\prime}\frac{(\bfx-\bfc^{*})}{\|\bfx-\bfc^{*}\|}\right| \leq 2 (\|\bfx\|+\|\bfc^{*}\|)\|\bfc_1-\bfc_2\|\leq M\delta_\epsilon
	\end{eqnarray*}
	by the multivariate mean value theorem where $\bfc^{*}=t\bfc_1+(1-t)\bfc_2$, $0<t<1$.
	The constant $M$ depends on the upper bound of $\|\bfx\|$ and $a$ only. Similarly, $\left|r_1^2-r_2^2\right|=(r_1+r_2)|r_1-r_2|<2\bar{r}^{*} \delta_\epsilon$.
	
	Let $\nu\coloneqq d/2-1$ and observe that
	\begin{eqnarray*}
		\frac{C_d\left(\|\tau_m\bfmu_m+r_2(\bfx-\bfc_2)/\sigma^2\|\right)  }{C_d\left(\|\tau_m\bfmu_m+r_1(\bfx-\bfc_1)/\sigma^2\|\right)}=
		\frac{ \|\tau_m \bfmu_m +r_1(\bfx-\bfc_1)/\sigma^2\|^{-\nu} I_{\nu}\left( \|\tau_m \bfmu_m +r_1 (\bfx-\bfc_1)/\sigma^2\| \right) }{\|\tau_m \bfmu_m +r_2(\bfx-\bfc_2)/\sigma^2\|^{-\nu} I_{\nu}\left( \|\tau_m \bfmu_m +r_2 (\bfx-\bfc_2)/\sigma^2\| \right)}.
	\end{eqnarray*}
	Note that $C_d^{-1}(\tau)$ is a continuous, strictly increasing function of $\tau$ (see \cite{MBFbound}). Further,
	\begin{align*}
	\|\tau_m \bfmu_m +r_1(\bfx-\bfc_1)/\sigma^2\| &\leq \|\tau_m \bfmu_m +r_2(\bfx-\bfc_2)/\sigma^2\| + r_2 \|\bfc_1 -\bfc_2\| + |r_1-r_2| \| \bfx - \bfc_1\| \\
	&\leq \|\tau_m \bfmu_m +r_2(\bfx-\bfc_2)/\sigma^2\| + \delta_\epsilon  r_2 + \delta_\epsilon \| \bfx - \bfc_1\|= g(\bfc_1,\bfc_2,r_2).
	\end{align*}
	Next note that $g(\cdot)$ is a continuous function on a compact set, and hence the image of $g(\cdot)$ is compact. Thus, $C_d(\cdot)$ is uniformly continuous on the image of $g(\cdot)$, and we can write
	$$\frac{C_d(\|\tau_m \bfmu_m  +r_2(\bfx-\bfc_2)/\sigma^2\|)  }{C_d(\|\tau_m  \bfmu_m  +r_1(\bfx-\bfc_1)/\sigma^2\|)}=1+\xi_\delta, $$
	where $\xi_\delta\rightarrow 0 $ as $\delta_\epsilon\rightarrow0$.
	Therefore, $h_\sigma(\bfx,\bta_1)/h_\sigma(\bfx,\bta_2) \leq\exp\left(M^{*}\delta_\epsilon\right)$ for some suitable constant $M^{*}$. Similarly, it can be shown that $h_\sigma(\bfx,\bta_1)/h_\sigma(\bfx,\bta_2)\geq \exp\left(-M^{*}\delta_\epsilon\right)$. Hence the above uniformly converges to 1 as $\delta_\epsilon \rightarrow 0$.
	
	Again, from Lemma \ref{lm:3}
	~$\exp\left(-U/\sigma^2 \right) \leq
	\left(2\pi\sigma^2\right)^{d/2} h_\sigma(\bfx,\bta_2) \leq \exp\left(U/\sigma^2 \right),$~
	where $U=\left(\|\bfx\|^2+a^2+\bar{r}^{*2}+2\tau_m \right)$.	
	Hence $h_\sigma(\bfx,\bta_2)$ is bounded away from zero and bounded above in $\bfx\in C$ and $\bta_2 \in E$. This completes the proof of equicontinuouity.
	
	The proof is complete if we show that $\sup\left\{\sum_j \pi_{j}^{(\bfc,r)} h_\sigma(\bfx,\bta): \bfx\in C, \bta\in E^c \right\}<k\varepsilon/4$. For that we just show $\sup\left\{ h_\sigma(\bfx,\bta): \bfx\in C, F \right\}<k\varepsilon/4$, where $F=\{(\bfc,r) : \|c\|>a; ~\mbox{or}~ r>\bar{r}^{*} \}$. As under $E$, $\bfpi$ is any point in $\Delta_M$, $E^c=\left(F\cup \{ \phi\}\right)$ and the result follows. 
	
	Let $\|\bfx\|<m^{*}$ if $\bfx\in C$. Also, we can choose $a$ such that whenever $\|\bfc\|>a$ then  $\|\bfx-\bfc\|\geq M_\varepsilon$.
	We split the space $\{(\bfx,\bta): \bfx\in C; (\bfc, r) \in F \}$ into two parts:
	\begin{eqnarray*}
		A=\{(\bfx,\bta): \left(\bfc,r\right)\in F \quad\mbox{and}\quad \left(\|\bfx-\bfc\|-r\right)^2 > -2\sigma^2\log \left(C_\sigma k \varepsilon/ 4\right) \};\\
		B=\{(\bfx,\bta):  \left(\bfc,r\right)\in F \quad\mbox{and}\quad \left(\|\bfx-\bfc\|-r\right)^2 \leq -2\sigma^2\log \left(C_\sigma k\varepsilon/ 4 \right) \}.
	\end{eqnarray*}
	Here $C_\sigma=e^{-2\tau_{\max}}(2\pi\sigma)^{d/2}$. Further we consider $\varepsilon$ to be small enough so that $C_\sigma k\varepsilon<1$.
	Observe that
	$$ \sup_{(\bfx,\bta)\in A} h_\sigma(\bfx,\bta) \leq \frac{e^{2\tau_{\max}}}{(2\pi\sigma)^{d/2}}\exp\left\{ -\frac{1}{2\sigma^2} \inf_{(\bfx,\bta)\in A} \left(\|\bfx- \bfc\|-r\right)^2 \right\} < k\varepsilon/ 4.$$
	
	For $B$, first note that when $\bta\in E^c$, either $\|\bfc\|>a$ or $r>\bar{r}^{*}$, or both happen. Let $\tau_m \geq \|\tau_m \bfmu_m +r(\bfx-\bfc)/\sigma^2 \|$. Note that $C_d^{-1}(\tau_m)$ is a strictly increasing function of $\tau_m$. Therefore, $  C_d^{-1}(\tau_m) \geq C_d^{-1} \left(\|\tau_m \bfmu_m + r (\bfx-\bfc) /\sigma^2 \| \right),$ implying $C_d(\tau_m)/C_d\left(\|\tau_m \bfmu_m +r(\bfx-\bfc)/\sigma^2 \|\right)\leq 1.$ Further, as $\left( \bfc,r \right)\in F$, for fixed $\sigma^2$ and suitable choices of $a$ and $\bar{r}^*$, $\left( \|\bfx-\bfc\|^2+r^2 \right)/2\sigma^2 > -\log \left( C_\sigma k \varepsilon \right)$, $\sup_{(\bfx,\bta)\in A} h_\sigma(\bfx,\bta) \leq k\varepsilon$.
	
	Next consider $\|\tau_m \bfmu_m +r(\bfx-\bfc)/\sigma^2 \|>\tau_m $. By \citet[Equation (2.10)]{MBFbound} for $0<\tau_m <\|\tau_m \bfmu_m +r(\bfx-\bfc)/\sigma^2 \|$,
	\begin{align*}
	\frac{C_d(\tau_m)}{C_d(\|\tau_m \bfmu_m +r(\bfx-\bfc)/\sigma^2 \|)}=\frac{\tau_m^{d/2-1}I_{d/2-1}\left(\|\tau_m \bfmu_m +r(\bfx-\bfc)/\sigma^2 \|\right)}{\|\tau_m \bfmu_m +r(\bfx-\bfc)/\sigma^2 \|^{d/2-1}I_{d/2-1}(\tau_m )}\hspace{1 in} \\
\hspace{.7 in}	<\left(\frac{\tau_m +d/2-1}{\|\tau_m \bfmu_m +r(\bfx-\bfc)/\sigma^2 \|+d/2-1} \right)^{d/2-1}\exp\left\{ \|\tau_m \bfmu_m +r(\bfx-\bfc)/\sigma^2 \|-\tau_m  \right\}.
	\end{align*}
	Therefore, $ \sup_{(\bfx,\bta)\in B}h_\sigma(\bfx,\bta)$ is no bigger than
	\begin{eqnarray*}
		&&\frac{(\tau_m +d/2-1)^{d/2}}{C_{\sigma} \exp(2\tau_m )} \sup_{(\bfx,\bta)\in B} \left\{ \|\tau_m \bfmu_m +r(\bfx-\bfc)/\sigma^2 \|+d/2-1\right\}^{-(d/2-1)}\\
		&& \hspace{3 in} \exp\left\{ -\frac{1}{2\sigma^2} \inf_{(\bfx,\bta)\in B} (\|\bfx-\bfc\|-r)^2 \right\} \\
		&& \leq \frac{(\tau_m +d/2-1)^{d/2}}{C_{\sigma} \exp(2\tau_m )}  \sup_{(\bfx,\bta)\in B} \left\{ \|\tau_m \bfmu_m +r(\bfx-\bfc)/\sigma^2 \|+d/2-1\right\}^{-(d/2-1)}.
	\end{eqnarray*}
	For sufficiently large $M_\varepsilon$ and $\bar{r}^{*}$ we can write the bracketed portion in the last expression as $ \|\tau_m \bfmu_m +r(\bfx-\bfc)/\sigma^2 \|+d/2-1\geq c^{*} r\|\bfx-\bfc\|/\sigma^2$, for some suitable constant $c^{*}$. Note that under $B$, $\left(r-\|\bfx-\bfc\|\right)^2<-2\sigma^2\log\left(C_{\sigma} k\varepsilon\right)=\delta_\varepsilon$, say, implying $r\|\bfx-\bfc\|> \left(r^2+ \|\bfx-\bfc\|^2-\delta_\varepsilon \right)/2 $. Therefore, we can expand $E$ by choosing $a$ and $\bar{r}^{*}$ such that the last expression is less than $k\varepsilon$ for $d >2$.

	Finally consider the case with $d=2$. Fix a small $\epsilon=1/4$. By \cite[Theorem 2.2]{MBFbound} 
	$I_0(\tau)/I_{-0.5+\epsilon}(\tau)$ is a strictly decreasing function of $x$. So for any $\tau>1$,
	\begin{eqnarray*}
		\frac{I_0(\tau)}{I_{-0.5+\epsilon}(\tau)} < \frac{I_0(1)}{I_{-0.5+\epsilon}(1)} \approx
		\frac{1.2661}{1.3178}=c^{*} ~
		~ \implies I_0(\tau)\leq c^{*} I_{-0.5+\epsilon}(\tau).
	\end{eqnarray*}
	Therefore, for $d=2$,
	\begin{eqnarray*}
		\frac{C_d(\tau_m)}{C_d\left(\|\tau_m \bfmu_m +r(\bfx-\bfc)/\sigma^2 \|\right)} &=& \frac{I_{0}\left(\|\tau_m \bfmu_m +r(\bfx-\bfc)/\sigma^2 \|\right)}{I_{0}(\tau_m)}\\
		&\leq&  M\frac{I_{-0.5+\epsilon}\left(\|\tau_m \bfmu_m +r(\bfx-\bfc)/\sigma^2 \|\right)}{I_{-0.5+\epsilon}(\tau_m )},
	\end{eqnarray*}
	where $M=c^{*} I_{-0.5+\epsilon}(\tau_m)/I_{0}(\tau_m)>0$ as without loss of generality for $(\bfc,r)\in B$, we can assume that $\|\tau_m\bfmu_m+r(\bfx-\bfc)/\sigma^2\|>1$. Next by \cite{joshi_bissu_1991}, as $\|\tau_m\bfmu_m+r(\bfx-\bfc)/\sigma^2 \|>\tau_m$, $I_{-0.5+\epsilon}\left\{\|\tau_m \bfmu_m +r(\bfx-\bfc)/\sigma^2 \|\right\}/I_{-0.5+\epsilon}(\tau_m)$ is no bigger than
	\begin{eqnarray*}
		\left\{ \frac{\tau_m}{\|\tau_m \bfmu_m +r(\bfx-\bfc)/\sigma^2 \|} \right\}^{0.5-\epsilon}\exp\left(\|\tau_m \bfmu_m +r(\bfx-\bfc)/\sigma^2 \|-\tau_m \right) \\
		\leq  \left\{ \frac{\tau_m }{\|\tau_m\bfmu_m+r(\bfx-\bfc)/\sigma^2 \|} \right\}^{0.5-\epsilon}\exp\left(\frac{r}{\sigma^2}\|\bfx-\bfc\|\right).
	\end{eqnarray*}
	Thus $ \sup_{(\bfx,\bta)\in B}h_\sigma(\bfx,\bta)$ is no bigger than
	\begin{eqnarray*}
		M^{*} \sup_{(\bfx,\bta)\in B}  \left\{ \frac{\tau_m}{\|\tau_m \bfmu_m +r(\bfx-\bfc)/\sigma^2 \|} \right\}^{0.5-\epsilon} \exp\left\{ -\frac{1}{2\sigma^2} \inf_{(\bfx,\bta)\in B} (\|\bfx-\bfc\|-r)^2 \right\} \\
		\leq M^{*} \sup_{(\bfx,\bta)\in B}  \left\{ \frac{\tau_m }{\|\tau_m \bfmu_m +r(\bfx-\bfc)/\sigma^2 \|} \right\}^{0.5-\epsilon},
	\end{eqnarray*}
	for an appropriate constant $M^{*}$.
	As $\epsilon=1/4$, we can enlarge $E$ to show that last expression is less than $k\varepsilon/4$ for all $\tau_m$, $m=1,\ldots, M$. 
	This completes the proof.
\end{proof}

	\bibliographystyle{apalike} 
	\bibliography{ref}

\end{document}